\documentclass[prd,reprint,nofootinbib,floatfix,
showkeys,aps,preprintnumbers,superscriptaddress,
numbers,square,comma,sort&compress,merge]{revtex4-1}
\usepackage[utf8]{inputenc}
\usepackage[spanish]{babel}
\usepackage{graphicx}
\usepackage{epsf,amsmath,bbold,amsfonts,stmaryrd}
\usepackage{mathrsfs}
\usepackage{appendix}
\usepackage{amssymb}
\usepackage{float}
\usepackage{color} 
\usepackage[colorlinks]{hyperref}
\hypersetup{pageanchor=false,citecolor=blue,
urlcolor=blue,linkcolor=blue}
\usepackage{indentfirst}
\usepackage[]{natbib}
\usepackage[normalem]{ulem}

\def\a{\alpha}
\def\b{\beta}

\def\g{\gamma}

\def\G{\Gamma}

\def\s{\sigma}

\def\bea{\begin{eqnarray}}
\def\eea{\end{eqnarray}}
\def\ba{\begin{array}}
\def\ea{\end{array}}
\def\bc{\begin{center}}
\def\ec{\end{center}}
\def\bl{\begin{flushleft}}
\def\el{\end{flushleft}}
\def\br{\begin{flushright}}
\def\er{\end{flushright}}
\def\bi{\begin{itemize}}
\def\ei{\end{itemize}}

\def\bt{\begin{tabular}}
\def\et{\end{tabular}}
\newtheorem{question}{Question}
\def\bq{\begin{question}}
\def\eq{\end{question}}
\newtheorem{definition}{Def}
\def\bd{\begin{definition}}
\def\ed{\end{definition}}
\newtheorem{answer}{Answer}
\def\ban{\begin{answer}}
\def\ean{\end{answer}}
\newtheorem{possibleanswer}{Possible answer}
\def\bpa{\begin{possibleanswer}\normalfont}
\def\epa{\end{possibleanswer}}
\newtheorem{theorem}{Theorem}
\def\bth{\begin{theorem}}
\def\eth{\end{theorem}}

\newcommand{\be}[1]{\begin{equation} \centering \label{#1}}
\newcommand{\ee}{\end{equation}}

\begin{document}
\title{One residue to rule them all: \\ 
Electroweak symmetry breaking, inflation and field-space geometry}

\author{Georgios K.~Karananas}
\email{georgios.karananas@physik.uni-muenchen.de}
\affiliation{Arnold Sommerfeld Center, 
Ludwig-Maximilians-Universit\"at M\"unchen,  Theresienstra{\ss}e 37,
80333 M\"unchen, Germany} 
\author{Marco Michel}
	\email{michel@mpp.mpg.de}
\affiliation{Arnold Sommerfeld Center, 
Ludwig-Maximilians-Universit\"at M\"unchen,  Theresienstra{\ss}e 37, 
80333 M\"unchen, Germany} 
	\affiliation{ Max-Planck-Institut f\"ur Physik, 
F\"ohringer Ring 6, 80805  M\"unchen, Germany}
\author{Javier Rubio}
 	\email{javier.rubio@tecnico.ulisboa.pt}
\affiliation{Department of Physics and Helsinki Institute of Physics,  
PL 64, FI-00014 University of Helsinki, Finland}
\affiliation{Centro de Astrofisica e Gravita\c c\~ao-CENTRA,
Departamento de Fisica, Instituto Superior T{\'e}cnico-IST,
Universidade de Lisboa-UL, Avenida Rovisco Pais 1, 1049-001, Lisboa,
Portugal}

\begin{abstract}

We point out that the successful generation of the electroweak scale
via gravitational instanton configurations in certain scalar-tensor
theories can be viewed as the aftermath of a simple requirement: the
existence of a quadratic pole with a sufficiently small residue in the
Einstein-frame kinetic term for the Higgs field. In some cases, the
inflationary dynamics may also be controlled by this residue and
therefore related to the Fermi-to-Planck mass ratio, up to possible
uncertainties associated with the instanton regularization.  We
present here a unified framework for this hierarchy generation
mechanism, showing that the aforementioned residue can be associated
with the curvature of the Einstein-frame target manifold in models
displaying spontaneous breaking of dilatations. Our findings are
illustrated through examples previously considered in the literature.

\end{abstract}

\preprint{ LMU-ASC 29/20}
\preprint{MPP-2020-102}
\preprint{HIP-2020-19/TH}
\maketitle

\section{Introduction and summary}

The seminal discovery of the Higgs field at the LHC has left us with a
perfect Standard Model (SM) of particle physics potentially valid up
to energies well above the Planck scale $M_P=2.48 \times 10^{18}$
GeV. At the same time, it left unsolved one of the most mysterious
puzzles in particle physics: the so-called hierarchy
problem.\footnote{For another point of view, see
e.g. Ref.~\cite{Senjanovic:2020pnq}.} This has two facets. The first
one is the extreme sensitivity of the Higgs mass to whatever happens
above the electroweak scale.  Several ways of overpassing this
difficulty have been proposed in the literature. One of them is to
require new physics to appear around the TeV scale (e.g. low-energy
supersymmetry, technicolor/composite Higgs, large extra dimensions,
see for instance
Refs.~\cite{Feng:2013pwa,Bellazzini:2014yua,ArkaniHamed:1998rs}). Another
option is to postulate a dynamical relaxation mechanism, like the
cosmological attractor
scenario~\cite{Dvali:2003br,Dvali:2004tma,Dvali:2019mhn,Michel:2019vuv}
or its recent variants and
generalizations~\cite{Graham:2015cka,Kaplan:2015fuy}. Alternatively,
one could require the absence of additional particle states all the
way up till the Planck
scale~\cite{Vissani:1997ys,Shaposhnikov:2007nj,Shaposhnikov:2008xi,
Farina:2013mla,Karananas:2017mxm,Shaposhnikov:2018nnm}. Of course, the
long-standing question of what happens around and beyond that point
still remains. A priori, it is conceivable that quantum gravity
corrections may either turn out to be negligibly small, or take care
of the problem completely~\cite{Wetterich:2011aa, Wetterich:2016uxm}.
In addition, it might be the case that the fundamental gravitational
degrees of freedom above $M_P$ are black holes~\cite{Dvali:2010bf},
being their influence on low-energy physics exponentially suppressed
at least by a Boltzmann factor proportional to the
entropy~\cite{Dvali:2010ue}. We will content here with assuming that,
if such contributions are present to start with, the theory is
liberated from them in one way or another. This leaves us with the
second facet of the hierarchy problem: the origin of the 16 orders of
magnitude difference between the electroweak and the Planck scale.

Non-perturbative effects constitute a natural tool for obtaining
``small numbers,'' especially in models with negligible perturbative
corrections. This possibility has been advocated in certain
scalar-tensor theories~\cite{Shaposhnikov:2018xkv} and generalized to
scale-invariant models where the Planck mass is generated by the
spontaneous breaking of
dilatations~\cite{Shaposhnikov:2018jag,Shkerin:2019mmu}, showing
explicitly that a second scale can be dynamically generated by an
instanton configuration. This idea was recently extended to the
Palatini formulation of gravity~\cite{Shaposhnikov:2020geh}.

In this short paper we generalize the findings of
Refs.~\cite{Shaposhnikov:2018jag,Shaposhnikov:2018xkv,
  Shkerin:2019mmu,Shaposhnikov:2020geh},
isolating the fundamental ingredients for successfully generating the
electroweak scale via instanton effects. In particular, we argue that:

\begin{enumerate}
\item Any scenario able to bring the (conformal) SM scalar sector at
large Higgs values to the approximate form
\be{eq:th}
\frac{\mathscr L}{\sqrt{g}}\approx \frac{M_P^2}{2} R - \frac{1}{2}
\frac{M_{P}^2}{\vert \kappa_c \vert} \frac{(\partial
  \Theta)^2}{\Theta^2} - V_0\ , \nonumber 
\ee
with $g$ the metric determinant, $R$ the scalar curvature,
$\Theta\propto h^{-1}$, $h$ the Higgs field in the unitary gauge and
$V_0$ an approximately constant potential, will be able to generate a
large hierarchy among the electroweak and the Planck scale for
sufficiently large values of the inverse residue $\vert
\kappa_c\vert$.
\item Provided that the scale $V_0$ is compatible with the COBE
normalization~\cite{Akrami:2018odb}, the inverse residue $\vert
\kappa_c\vert$ controls also the inflationary
observables. Consequently, if the above splitting mechanism is
operative, inflation is intimately related to the electroweak symmetry
breaking, making a priori possible to infer the value of the Fermi
scale from CMB
observations~\cite{Shaposhnikov:2020geh,Shaposhnikov:2020fdv}.
    \item The above reasoning holds true irrespectively of the nature
of the gravitational interaction (metric, Palatini, Einstein-Cartan
\ldots). The difference boils down to the pole structure of the Higgs
kinetic term in the large field regime and, in particular, to the
value of the inverse residue $\vert \kappa_c\vert$.
    \item When single-field models involving the Higgs field are
embedded into a fully scale-invariant two-field framework, $\vert
\kappa_c\vert $ becomes the curvature of the target manifold at large
field values.
    All previous considerations continue to apply for a large class of
models displaying a maximally symmetric Einstein-frame kinetic
sector. However, unlike the single-field case, the requirement of
scale/conformal symmetry is not enough for the successful
implementation of the proposed hierarchy generation in biscalar
theories. It is also important that no mass term for the Higgs field
is (classically) generated by the spontaneous breaking of scale
invariance.
\end{enumerate}
Having identified the essence of the mechanism leading to the
generation of well separated scales, the approach presented here opens
up a new avenue for model building by rephrasing the usual hierarchy
problem as a question about the field-space geometry. Among other
applications, this could lead to interesting synergies with
$\alpha$-attractors~\cite{Kallosh:2013hoa,Galante:2014ifa,Kallosh:2015zsa,
  Artymowski:2016pjz,Fumagalli:2016sof} 
and superconformal field
theories~\cite{Cremmer:1977tt,Bergshoeff:1980is,deRoo:1984zyh,Ferrara:2012ui}.

\section{Higgs' pole structure}

The SM Lagrangian acquires (classical) conformal invariance when the
electroweak scale is set to zero. As can be explicitly seen in the
minimal subtraction scheme~\cite{tHooft:1973mfk}, this implies that no
counterterm is needed to renormalize the Higgs mass, which becomes
then computable in terms of other SM
parameters~\cite{Weinberg:1973am,Weinberg:1976pe,Linde:1977mm}. Interestingly
enough, this appealing property remains true even in the presence on a
non-minimal coupling to gravity~\cite{Shaposhnikov:2020geh}. Having
this in mind, we consider the following non-trivial extension of the
conformal SM Higgs sector,
\be{eq:higgs_inflation}
\frac{\mathscr L}{\sqrt{g}}=\frac{M_P^2+\xi_h h^2}{2}\,g^{\mu\nu}
R_{\mu\nu}(\G)-\frac{1}{2}(\partial h)^2- \frac{\lambda}{4} h^4 \ . 
\ee
Here $\xi_h>0$ controls the strength of the Higgs coupling to gravity
and $\lambda$ is the field's quartic self-interaction. Note that for
the sake of generality, we have not identified the connection
determining the Ricci tensor $R_{\mu\nu}(\G)$ with the Levi-Civita
one. Nevertheless, we will assume it to be symmetric ($\G^\a_{\b\g}=
\G^\a_{\g\b}$) in what follows, such that the considered set of
theories are torsionless (for non-vanishing torsion scenarios, see
e.g.~Ref.~\cite{Rasanen:2018ihz}).

In order to simplify the analysis, it is convenient to get rid of the
non-minimal coupling to gravity by moving to the so-called Einstein
frame. This is achieved by considering a Weyl rescaling of the metric
$g_{\mu\nu}\to \omega^2 g_{\mu\nu}$, with conformal factor $\omega^{2}
= (M_{P}^2+\xi_h h^2)/M_{P}^2$. After some trivial algebra, we get
\be{eq:L_EF}
\frac{\mathscr L}{\sqrt{g}}=\frac{M_P^2}{2} R - \frac{1}{2} \gamma(h)
(\partial h)^2 - \frac{\lambda M_{P}^4 h^4}{4(M_{P}^2+\xi_h h^2)^2}\ , 
\ee
with 
\be{eq:curv_single_field}
\gamma(h) = \frac{M_{P}^2}{M_{P}^2+\xi_h h^2}\left(1+
  \frac{6\,\a\,\xi_h^2 h^2}{M_{P}^2+\xi_h h^2}\right) \ , 
\ee
and $\alpha=0$ or $1$ for the Palatini or metric formulations,
respectively. The essential effect of the Weyl transformation is to
transfer the non-linearities associated with the Higgs non-minimal
coupling to the scalar sector of the theory.  While in the Palatini
formulation the connection---and consequently the Ricci tensor---is
inert under Weyl rescalings, this is not the case in the metric
scenario, where the dependence of the Levi-Civita connection on the
metric leads to an additional contribution in
Eq.~(\ref{eq:curv_single_field}).  Introducing a variable
$\Theta=M_{P}/\sqrt{M_{P}^2+\xi_h h^2}$, we find that for field values
relevant for inflation ($h\gg M_{P}/\sqrt{\xi_h}$, or equivalently
$\Theta\ll 1$),\footnote{A field redefinition $\Theta\propto h^{-1}$
is convenient to highlight similarities with the two-field scenarios
considered in Section~\ref{sec:two-field}, where the inflationary
region is restricted to a compact field range.} the
Lagrangian~(\ref{eq:L_EF}) can be well approximated by
\be{eq:HI_lagr_theta}
\frac{\mathscr L}{\sqrt{g}}\approx\frac{M_P^2}{2} R -
\frac{M_P^2}{2|\kappa_c|} \frac{(\partial\Theta)^2}{\Theta^2} -
\frac{\lambda M_P^4}{4\xi_h^2}(1-\Theta^2)^2  \ , 
\ee
with 
\be{eq:kap_c}
\kappa_c \equiv-\frac{\xi_h}{1+6\,\alpha\,\xi_h}\ .
\ee
The pole structure in the above allows for inflation, while making the
inflationary observables almost insensitive to the details of the
potential~\cite{Karananas:2016kyt,Karananas:2016grc,Casas:2017wjh,Galante:2014ifa} 
(for a review, see Ref.~\cite{Rubio:2018ogq}). The spectral tilt and
tensor-to-scalar ratio
\be{eq:Obs_SF}
n_s\simeq 1-\frac{2}{N_*}\ , \hspace{15mm} r = \frac{2}{ \vert
  \kappa_c\vert  N_*^2}\ , 
\ee
depend only on the number of e-folds of inflation $N_*$, dictated by
the post-inflationary
dynamics~\cite{GarciaBellido:2008ab,Bezrukov:2008ut,Repond:2016sol,Rubio:2019ypq}
and the inverse of the residue at the inflationary pole at $\Theta=0$,
namely
\be{eq:residueHI}
\vert \kappa_c\vert \approx\begin{cases} \xi_h
  \hspace{12mm}\text{Palatini} \ ,\\ 
1/6 \hspace{10mm} \text{metric} \ ,
\end{cases}
\ee
where in the last step we have taken into account the well-known
restriction $\xi_h\gg 1$ needed to generate the right amplitude of
primordial density
perturbations~\cite{Rubio:2019ypq,Shaposhnikov:2020fdv}. Note that
this result unifies those in
Refs.~\cite{Bezrukov:2007ep,Bauer:2008zj}, see also
Refs. \cite{Bezrukov:2017dyv,
  Fumagalli:2016lls,Rasanen:2017ivk}.

Interestingly, the inverse residue~(\ref{eq:residueHI}) controls the
Higgs vacuum expectation value (vev), \be{eq:pathI} \langle h \rangle
\sim \int \mathcal D \varphi \,\, h \,\, e ^ {-S_E} \ , \ee with the
path integral taken over all fields (including the metric) and $S_E$
the Euclidean action of the theory. After canonically normalizing the field as $\Theta =
\exp(-\sqrt{|\kappa_c|}\, \theta/M_P)$, this equation becomes roughly
\be{eq:pathI2}
\langle h\rangle\sim\dfrac{M_P}{\sqrt{\xi_h}}\int D\varphi \, J\,
e^{-W} \ , 
\ee
with $J$ the Jacobian of the transformation and
\be{eq:W}
W = S_E - \sqrt{|\kappa_c|}\frac{\theta(0)}{M_P} \ .
\ee
Note that without loss of generality, we have taken the (instantaneous
and localized) source of the scalar field to be at the origin of
coordinates.

Assuming the dominant contribution to Eq.~(\ref{eq:pathI2}) to be
determined by the extrema of $W$ after regulating the theory with an
appropriate higher-dimensional operator (cf.~Appendix for details),
the vev in the saddle-point approximation
becomes~\cite{Shaposhnikov:2018xkv,Shaposhnikov:2018jag,
  Shkerin:2019mmu,Shaposhnikov:2020geh}
\be{eq:vev_h} 
\langle h \rangle \sim \frac{M_{P}}{\sqrt{\xi_h}}e^{-{\cal
    W}(|\kappa_c|)} \ , 
\ee
with ${\cal W}(|\kappa_c|)$ a function of the inverse residue $\vert
\kappa_c\vert$, whose precise expression is irrelevant for the present
discussion. It suffices to point out that it remains finite and that
the bigger the inverse residue $|\kappa_c|$, the larger the
exponential suppression.  An accurate result accounting for the
regularization of the gravitational instanton can be obtained by
numerically solving the system of equations \eqref{eq:A1} in the
Appendix, as done for instance in
Refs.~\cite{Shaposhnikov:2018xkv,Shaposhnikov:2018jag,
  Shkerin:2019mmu,Shaposhnikov:2020geh}.

A simple inspection of Eq.~(\ref{eq:kap_c}) reveals that the value of
$\vert \kappa_c\vert$ in the metric formulation ($\a=1$) is restricted
to an ${\cal O}(1)$ range, $0<|\kappa_c|\le 1/6$, meaning that a
satisfactory splitting between the electroweak and Planck scales
cannot be obtained in the most ``vanilla''version of this scenario. In
order to reproduce the observed hierarchy, one must inevitably modify
the value of the inverse residue at high energies. This can be done in
two ways. The first one is to change explicitly the structure of the
kinetic sector, as done for instance in
Ref.~\cite{Shaposhnikov:2018xkv}. The second one is to work directly
in the Palatini formulation, where $|\kappa_c|\approx \xi_h$ and
${\cal W}(\vert \kappa_c\vert)\gg 1$.~This change of gravity paradigm
translates automatically into a larger exponential
suppression~\cite{Shaposhnikov:2020geh}.~Note, however, that, as seen
from the unifying glass advocated here, the distinction between these
two approaches loses importance. Indeed, when written in the original
frame~(\ref{eq:higgs_inflation}), the metric and Palatini formulations
differ only by an asymptotically scale-invariant higher-dimensional
operator $-3\xi^2 h^2(\partial h) ^2/(M_P^2+\xi_h h^2)$, as those
required for the self-consistency of the metric effective theory about
its cutoff scale~\cite{Bezrukov:2010jz,Bezrukov:2014ipa,Bezrukov:2017dyv}.\footnote{The
non-minimal coupling of the Higgs field to gravity makes this theory
perturbatively non-renormalizable, meaning that it should be
interpreted as an effective field theory to be complemented by a set
of higher-dimensional operators suppressed by a given cutoff scale. As
shown in Refs.~\cite{Bezrukov:2010jz,Bezrukov:2014ipa}, the
requirement of having a UV completion respecting the symmetries of the
original theory forbids the appearance of dangerous operators
potentially spoiling the flatness of the inflationary potential. The
minimal set of operators is induced by the theory itself as an
infinite tower of counterterms aimed to cancel the loop divergences
generated order by order in perturbation theory. As compared to the
tree-level action, these operators display a hierarchical structure
when written in the Einstein-frame, being exponentially suppressed at
large field values.  The \textit{collective} effect of all
higher-dimensional operators generated by the tree-level action
manifest only as localized threshold effects in the SM renormalization
group running~\cite{Bezrukov:2014ipa,Bezrukov:2017dyv}. Although these
are important for determining the precise value of the non-minimal
coupling to gravity~\cite{Bezrukov:2014ipa,Bezrukov:2017dyv}, they
play no role in the (qualitative) instanton dynamics if the associated
residue $\vert \kappa_c \vert $ stays sufficiently large. In
particular, a change in the self-coupling $\lambda$ by threshold
corrections will translate just into a shift of the inverse residue
$\vert \kappa_c\vert$.} From this point of view, the Palatini
formulation could be understood as a particular low-energy truncation
of the unknown ultraviolet completion of the metric theory, written in
a convenient set of variables. The fact that the different approaches
followed in Refs.~\cite{Shaposhnikov:2018xkv}
and~\cite{Shaposhnikov:2020geh} were able to generate the required
hierarchy illustrates that the precise choice of higher-dimensional
operators with up to two derivatives of the field to be included in
the metric scenario is actually not relevant. For instance, the
inclusion in the initial frame of any asymptotically scale-invariant
operator of the form $-3 \xi^2 h^{n}(\partial h) ^2/(M_P^2+\xi_h
h^2)^{n/2}$ with even $n\geq 2$ would produce the same effect as the
operator differentiating metric and Palatini formulations.  The most
relevant ingredient, up to uncertainties associated with the instanton
regularization~\cite{Shaposhnikov:2018xkv,Shaposhnikov:2018jag,
Shkerin:2019mmu,Shaposhnikov:2020geh}, is the effective value of the
inverse residue $\vert \kappa_c\vert$ at large field values. Provided
this is large, the splitting of scales is guaranteed to take place.

\section{The geometrical picture}
\label{sec:two-field}

Having understood the key role of the Higgs inverse residue $\vert
\kappa_c\vert$ in the inflationary observables and the hierarchy
between the electroweak and Planck scales, we will show now that it
also has a nice geometrical interpretation.

As a warm up, let us extend the Lagrangian \eqref{eq:higgs_inflation}
by a dilaton field $\chi$, namely $\mathscr L' = \mathscr L + \mathscr
L _\chi$, with $\mathscr L_\chi/\sqrt{g}=-\frac{1}{2}(\partial
\chi)^2$. After performing the same Weyl rescaling we did in the
previous section, the kinetic sector of the theory boils down to a
non-linear $\sigma$-model,
\be{}
\frac{\mathscr L '}{\sqrt{g}} \supset -\frac{1}{2}\g(h) \Big((\partial
h)^2 + (\partial \chi)^2\Big) \ , 
\ee
with Gaussian  curvature 
\be{eq:curvat}
\mathcal K = \frac12 \frac{\g'^2(h)-\g(h)\g''(h)}{\g^3(h)} \ .
\ee
Here $\g(h)$ is given by Eq.~(\ref{eq:curv_single_field}) and the
primes denote differentiation with respect to $h$. At large field
values, the Gaussian curvature~(\ref{eq:curvat}) becomes approximately
constant and coincides, in Planckian units, with the quantity
$\kappa_c$ in Eq.~(\ref{eq:kap_c}),
\be{}
 M_{P}^{2} \, \mathcal K \approx \kappa_c \ .
\ee
In other words, the residue of the Higgs kinetic pole is nothing else
than the dimensionless curvature of the Einstein-frame kinetic
manifold. Being a two-dimensional field space, this quantity
completely describes the geometry.

Now that we have established the geometrical meaning of $\kappa_c$, we
can go further and show that the results found in the single field
case hold true also in more general biscalar-tensor theories. Although
our considerations will be applicable well beyond a particular
scenario, we will focus here on the~\textit{Higgs-Dilaton model}~as a
proof of
concept~\cite{Shaposhnikov:2008xb,GarciaBellido:2011de,Bezrukov:2012hx,
GarciaBellido:2012zu,Rubio:2014wta,Casas:2017wjh,Casas:2018fum,
Herrero-Valea:2019hde} (for a comprehensive overview, see
Ref.~\cite{Rubio:2020zht}). As compared to the previous example, this
scale-invariant scenario includes also a positive non-minimal coupling
of the $\chi$ field to gravity, which effectively replaces the bare
Planck mass $M_P$ in Eq.~(\ref{eq:higgs_inflation}), i.e.
\be{eq:higgs-dilaton-lag}
\begin{aligned}
\frac{\mathscr L'}{\sqrt{g}}&= \frac{\xi_\chi \chi^2 + \xi_h h^2}{2}
g^{\mu\nu} R_{\mu\nu}(\G) \\
&\hspace{1.5cm}- \frac{1}{2} (\partial h)^2 - \frac{1}{2}
(\partial \chi)^2 - \frac{\lambda}{4} h^4 \ .
\end{aligned}
\ee
Some comments are in order at this point. The attentive reader will
have probably noticed that we have not included a scale (and
conformally) invariant $\alpha\chi^2 h ^2$ term, leading to a
non-vanishing expectation value for the Higgs field after the
spontaneous breaking of scale invariance, i.e. $\langle h \rangle
\propto \alpha \langle \chi \rangle $ with the parameter $\alpha$
tuned to a value $\alpha \sim 10^{-34} \xi_\chi$ in order to reproduce
the hierarchy of scales and $\xi_\chi \ll 1$ as required by
inflationary phenomenology~\cite{GarciaBellido:2011de}. The absence of
this operator is indeed crucial for the successful implementation of
the mechanism considered here and can be justified at different
levels. First, an explicit Higgs-Dilaton coupling can be excluded by
requiring the dilaton field $\chi$ to display an exact shift symmetry
in the matter sector, broken only mildly by gravitational
interactions. Interestingly, this would also forbid the inclusion of a
quartic dilaton self-interaction leading to a cosmological constant
term when moving to the Einstein frame.  Beyond symmetry restrictions,
the presence of a quartic mixing term could make the theory
ill-behaved in the ultraviolet domain \cite{Karananas:2016grc},
motivating also its exclusion on the basis of
self-consistency. Finally, a situation of this sort might arise
naturally in more ``exotic'' scenarios such as the Fishnet Conformal
Field Theory~\cite{Gurdogan:2015csr}, where it is indeed possible to
have spontaneous breaking of scale/conformal invariance without
generation of masses for the fields, at least in the large-$N$
limit~\cite{Karananas:2019fox}.

Provided that the mixing $\alpha\,\chi^2 h^2$ term is absent, either
because of one of the aforementioned reasons or by simply tuning its
coefficient to zero, the mass of the Higgs field is classically
zero. To guarantee that this is the case even when quantum corrections
are taken into account,\footnote{Note that if a non-vanishing mixing
term between the Higgs and the dilaton is perturbatively generated,
there is no sense in discussing non-perturbative ways of inducing the
Fermi scale, since these effects are always subdominant. In such a
case, a severe fine-tuning---completely analogous to the situation in
the SM---is needed for the electroweak-to-Planck scale hierarchy to be
reproduced. } the renormalization procedure must preserve the
symmetries of the classical theory at all orders of perturbation
theory, i.e. the scale invariance of the tree-level action and the
approximate shift symmetry of the field $\chi$. In the absence of
particle thresholds above the electroweak scale, this can be achieved
by employing a ``scale-invariant'' regularization
scheme~\cite{Englert:1976ep,Shaposhnikov:2008xi,Shaposhnikov:2008ar}. In
this setting, the 't Hooft-Veltman renormalization point $\mu$ in
dimensional regularization is replaced by a field dependent
renormalization point $\mu(h,\chi)$ accounting for the usual dimension
mismatch among bare and renormalized coupling constants. Different
choices of $\mu$ correspond to different theories. In particular, a
change of $\mu$ can always be compensated for by the inclusion of a
specific set of higher dimensional operators in the tree-level
action. For the sake of concreteness, we will assume the quantization
to be performed in the non-minimally coupled
frame~(\ref{eq:higgs-dilaton-lag}), according to the so-called
prescription I of Ref.~\cite{Bezrukov:2008ut}, where $\mu^2\sim
\xi_\chi \chi^2+\xi_h h^2$. Working in this context, the Higgs field
is completely desensitized from quantum corrections involving the
dilaton.  This remains true even when perturbative gravitational
corrections are taken into account. In particular, graviton loops
generate a contribution to the effective potential of the form
$\lambda^2 h^8/(\xi_\chi \chi^2+\xi_h h^2)^2$. In other words,
provided that the aforementioned assumptions hold, a Higgs mass term
cannot be generated at any order of perturbation theory in the $\alpha
\to 0$ limit; see
Refs.~\cite{Shaposhnikov:2008xi,Armillis:2013wya,Gretsch:2013ooa,Shaposhnikov:2018jag}.
The graviscalar instanton discussed in the previous section becomes
then a viable option to induce the electroweak scale. The detailed
analysis carried out in Ref.~\cite{Shaposhnikov:2018jag}, revealed
that it is indeed possible to generate a non-vanishing Higgs vev in
the Higgs-Dilaton
model via this mechanism. To illustrate this result in our language,
it is again convenient to perform the Weyl transformation $g_{\mu \nu}
\to \Omega^2 \,g_{\mu \nu}$ with conformal factor $\Omega^2=(\xi_\chi
\chi^2 + \xi_h h^2)/M_P^2$. This yields the Lagrangian density
\be{eq:LHD}
\frac{\mathscr L'}{\sqrt{g}} = \frac{M_P^2}{2} R - \frac{1}{2} g^{\mu
  \nu} \gamma_{ab} \partial_\mu \varphi^a \partial_\nu \varphi^b -
U(\varphi)\ , 
\ee
where we have organized the fields $\chi$ and $h$ into a vector
$\varphi^a = (\chi,h)$ with $a,b=1,2$ and defined the Einstein-frame
potential $U(\varphi) \equiv \Omega^{-4} V(\varphi)$. The field-space
metric in this expression
\be{eq:gammaab}
\begin{aligned}
\g_{ab} &= \frac{M_P^2}{\xi_\chi \chi^2+\xi_h h^2}\Bigg[
\begin{pmatrix}
1&0\\
0&1
\end{pmatrix} \\
&~~~~~~~~+\frac{6\,\a}{\xi_\chi \chi^2+\xi_h h^2}
\begin{pmatrix}
\xi_\chi^2\chi^2 & \xi_\chi \xi_h\chi  h \\
\xi_\chi \xi_h\chi  h & \xi_h^2h^2 
\end{pmatrix} \Bigg]\ ,
\end{aligned}
\ee
is a straightforward generalization of the coefficient $\gamma(h)$ in
Eq.~(\ref{eq:curv_single_field}). As before, the difference between
the metric and Palatini approaches is accounted for by the value of
$\a$. The kinetic sector in~(\ref{eq:LHD}) can be made diagonal by
introducing the
variables~\cite{Casas:2017wjh,Casas:2018fum,Almeida:2018oid}\,\footnote{Note
that $\Theta$ is restricted to the interval $\sigma \le \Theta \le
1$. In addition, we chose an ``exponential map'' in order to highlight
that $\Phi$ is the Goldstone boson associated with the non-linear
realization of scale symmetry--the dilaton.}
\bea
&\displaystyle\Theta = |\bar \kappa_c| \frac{(1+6\,\alpha\,
  \xi_\chi)\chi^2 + (1+6\,\alpha\, \xi_h)h^2}{\xi_\chi \chi^2 + \xi_h
  h^2}\,,&  \label{trans1} \\  
&\displaystyle e^{\frac{2 \sqrt{|\bar \kappa_c|} \Phi}{M_P}} =
\frac{|\kappa_c|}{|\kappa|} \frac{(1+6\,\alpha\, \xi_\chi)\chi^2 +
  (1+6\,\alpha\, \xi_h)h^2}{M_P^2}&\,,   \label{trans2} 
\eea
with  $\kappa_c$  defined in Eq.~(\ref{eq:kap_c}), and  
\be{}
\bar \kappa_c = \kappa_c \vert_{\xi_h\to \xi_\chi} \ ,\hspace{8mm}
\kappa = \kappa_c \left( 1- \frac{\xi_\chi}{\xi_h}\right)\ . 
\ee
In terms of these quantities, the Higgs-Dilaton Lagrangian
reads~\cite{Casas:2017wjh,Casas:2018fum,Almeida:2018oid} 
\be{eq:LagNewVariables} 
\frac{\mathscr L'}{\sqrt{\tilde{g}}} = \frac{M_P^2}{2} \tilde{R} -
\frac{K(\Theta)}{2} (\partial \Theta)^2 -\frac{\Theta}{2} (\partial
\Phi)^2 - U(\Theta)\ , 
\ee
with
\bea
\label{eq:kinetic_function}
\displaystyle K(\Theta) &=& \frac{M_P^2}{4\Theta }
\left(\frac{1}{\vert \kappa\vert  (\Theta-\sigma)} + \frac{1}{\vert
    \bar \kappa\vert (\Theta-1)} \right)\ ,\\ 
\displaystyle U(\Theta) &=& \frac{\lambda M_P^4}{4 \, \vert
  \bar\kappa\vert^2}\, (1-\Theta)^2\ , 
\eea
and
\be{}
 \bar \kappa= \kappa \vert_{\xi_h\to \xi_\chi} \ , \hspace{8mm}\sigma=
 \frac{\bar \kappa_c}{\kappa_c}\ . 
\ee
Note that the kinetic function $K$ in Eq.~(\ref{eq:kinetic_function})
has poles at $\Theta = 0$, $\sigma$ and $1$. The first two are
potentially explored during inflation, with $\sigma\propto \xi_\chi$
encoding the differences with the warm-up example above. The last one
is a ``Minkowski'' pole associated with the ground state of the
theory, namely $\langle h\rangle=0$, $\langle\chi \rangle \approx
M_{P}$, or equivalently, $\langle \Theta \rangle = 1$ and $\langle
\Phi \rangle = M_P/(2\sqrt{|\bar
\kappa_c|})\log{\left(1/(|\kappa|\sigma) \right)}$.

A geometrical interpretation of the inverse residues $\vert\kappa\vert
$ and $\vert \bar \kappa \vert$ can be obtained from the direct
computation of the Gaussian curvature of the target
manifold~(\ref{eq:gammaab}), namely
\be{}
\mathcal K\, = \frac{\kappa\, \bar\kappa}{M_P^2} \, \frac{
\kappa (\Theta-\sigma )^2+\bar\kappa(\Theta-1)^2 }{\left[\kappa
  (\Theta-\sigma)+\bar\kappa(\Theta-1 ) \right]^2}\ . 
\ee
As clearly illustrated by this expression, $\vert\kappa\vert$ and
$\vert \bar \kappa\vert $ coincide with the dimensionless curvature of
$\gamma_{ab}$ around their corresponding poles $\Theta=\sigma$ and
$\Theta=1$.

Focusing on the field values relevant for inflation and restricting
ourselves to the phenomenologically allowed limit $\xi_\chi\ll \xi_h$
($\sigma\ll \vert\kappa\vert \simeq
\vert\kappa_c\vert$)~\cite{GarciaBellido:2011de}, the kinetic
function~(\ref{eq:kinetic_function}) boils down to a quadratic pole
structure reminiscent of that found in the single field
scenario~(\ref{eq:HI_lagr_theta}), namely
\be{}
K(\Theta) \approx \frac{M_{P}^2}{4|\kappa_c|\Theta^2}  \ . 
\ee 
The residue of this pole controls again the inflationary observables
in the corresponding limit ~\cite{Casas:2017wjh,Almeida:2018oid}, as
well as the generation of the electroweak scale,
cf.~Eq.\eqref{eq:vev_h} and the Appendix. As in the singe field case,
the non-perturbative generation of the observed hierarchy
$\langle{h}\rangle \ll M_{P}$ cannot be achieved in the metric
scenario without appropriately modifying the kinetic sector of the
theory. This is again not the case in the Palatini formulation, where
the large value of the inverse residue $\vert \kappa_c\vert$ allows to
obtain the desired mass splitting.

The above results can be straightforwardly extended to a more general
class of scale-invariant models constructed on the basis of
volume-preserving
diffeomorphisms~\cite{Blas:2011ac,Karananas:2016grc}, provided that
they display the same ground state and a maximally symmetric
Einstein-frame kinetic manifold at large field values. As shown in
Refs.~\cite{Karananas:2016kyt,Casas:2018fum}, this additional symmetry
of the target manifold restricts the leading kinetic pole structure of
such models to be quadratic.  Whether the considerations presented in
this paper remain valid outside this equivalence class is still an
open issue. In particular, if present, higher-order poles should be
sufficiently suppressed---in a way similar to
Ref.~\cite{Broy:2015qna}---to guarantee the existence of a source term
selecting the graviscalar instanton solution advocated in
Refs.~\cite{Shaposhnikov:2018jag,Shaposhnikov:2018xkv,Shkerin:2019mmu,Shaposhnikov:2020geh}. Since
we do not aim to provide here an exhaustive classification of
plausible scenarios, but rather a guiding principle unifying existing
models in the literature, we postpone the study of this issue for the
future.

\section*{Acknowledgements}

We thank Gia Dvali and Goran Senjanovi\'c for general discussions on
the hierarchy problem and Mikhail Shaposhnikov, Andrey Shkerin, Inar
Timiryasov and Sebastian Zell for discussions and important comments
on the paper. JR acknowledges the support of the Fundação para a
Ciência e a Tecnologia (Portugal) through the CEEC-IND programme,
grant CEECIND/01091/2018.

\section*{Appendix: Instanton computation} 
\label{appendixA1}

We present here some details on the graviscalar instanton solution and
its relation to the inverse residue.  It should be clearly stated that
the following analytical estimates should be taken with a grain of
salt, being necessary a numerical treatment in order to extract
quantitative results. In this regard, we refer the reader to
Refs.~\cite{Shaposhnikov:2018xkv,Shaposhnikov:2020geh}.

\textit{Single field case---}Our starting point is Eq.~(\ref{eq:W}),
with the Euclidean action $S_E$ supplemented by a higher-dimensional
Einstein-frame operator with the approximate form $\beta\,
(\partial\theta)^6/M_{P}^8$ in the high-energy regime and taming an
unphysical field divergence at the origin.\footnote{In the frame
\eqref{eq:higgs_inflation}, this operator reads $\beta_0 \, (\partial
h)^6/(M_P^2+\xi_h h^2)^4$ with $\beta_0=\beta\,(\xi_h/\vert
\kappa_c\vert)^3$.}  We emphasize that this term is chosen here for
illustration purposes only, being the considerations below also valid
for more general higher-order operators~\cite{Shaposhnikov:2018jag}.
In particular, other higher-dimensional operators involving more than
two derivatives could be also considered. The exact form of these
terms is determined by the unknown UV completion of the theory and
cannot be addressed from the bottom-up approach presented
here. Different choices modify the precise way the instanton is
regularized at its core, leading to a different dependence of the
instanton action---and consequently of the scale splitting---on the
operators' coefficients.

As preferred by the symmetries of point-like sources, we will assume
the classical instanton configuration to be $O(4)$-symmetric, such
that the associated Euclidean metric takes the form ~$ds^2 = f^2(r)
dr^2 + r^2 d\Omega_3$\,,~with $r$ the radial component and $d\Omega_3$
the line element of the unit 3-sphere. Neglecting the contributions
from the potential and its derivative
\cite{Shaposhnikov:2018xkv,Shaposhnikov:2020geh},~the $rr$-component
of the Einstein equations and the scalar field equation of motion can
be written as
\be{eq:A1}
\begin{aligned}
&\frac{3-3\, f^2}{{\tilde r}^2} = \frac{1}{2} (\partial_{\tilde r}\,
\tilde \theta )^2 + \frac{5\,\beta}{f^4}(\partial_{\tilde r}\,
\tilde\theta)^6 
\ , \\
& \partial_{\tilde r}\,\tilde \theta +\frac{6\,\beta}{f^4}
(\partial_{\tilde r}\,\tilde \theta )^5 =
-\frac{\sqrt{|\kappa_c|}}{2\pi^2}\frac{f}{\tilde r^3}\ , 
\end{aligned}
\ee
with ${\tilde r}= r\, M_P$ and $\tilde \theta = \theta/M_P$
appropriate dimensionless variables. These equations are supplemented
by the flat boundary conditions $f(\tilde r)\sim 1$, $\tilde
\theta(\tilde r)\sim 0$ at $\tilde r\to \infty$.

We will be mainly interested in the behavior of the
system~(\ref{eq:A1}) in the core of the instanton, located at
distances $\tilde r \lesssim \tilde r_0\equiv
(\beta\,|\kappa_c|^2)^{1/12}$, where the higher-dimensional operator
$\beta (\partial\theta)^6/M_{P}^8$ dominates. The reason we focus on
this particular regime is because its dynamics is what influences the
instanton the most in the considered example, as shown in details in
Refs.~\cite{Shaposhnikov:2018xkv,Shaposhnikov:2018jag}. There, the
instanton solution asymptotes, up to order-one numerical
contributions, to
\be{eq:A2}
\begin{aligned}
\bar \theta&\sim -\log {\tilde r_0} \sim
-\log\left(\beta\,|\kappa_c|^2\right)\ ,\\ 
f &\sim  {\tilde r_0}^{4/5}\,\beta^{1/10}\,|\kappa_c|^{-3/10} \sim
\left(\beta^{-1}|\kappa_c|\right)^{-1/6} \ .  
\end{aligned}
\ee
Using these expressions one can easily show that the Euclidean action
evaluates to $S_E\sim \sqrt{|\kappa_c|}$, which in turn translates
into a large exponent ${\cal W}(|\kappa_c|)\sim
\sqrt{|\kappa_c\vert}\left(\log(\beta\,|\kappa_c|^2)-\mathcal
O(1)\right)$ in Eq.~(\ref{eq:vev_h}) and a small Higgs vev for
a sufficiently large inverse residue $\vert \kappa_c\vert$.

\emph{Instanton in biscalar case}---We turn now to the generalization
of the above result to the two-field case; see also
Ref.~\cite{Shaposhnikov:2018jag}. In particular, we are interested in
evaluating the action $W$ on the graviscalar instanton solution for
the Higgs-Dilaton scenario. The first step is to neglect the
low-energy pole at $\Theta = 1$ in Eq.~(\ref{eq:LagNewVariables}),
such that the target manifold becomes maximally symmetric. In
the limit $\s\ll|\kappa|\approx |\kappa_c|$, the ``temporal''
component of the Einstein equations and the equations of motion for
the scalar fields become respectively
\be{}
\begin{aligned}
& \frac{3-3 f^2}{{\tilde r}^2} =\frac{\Theta}{2}\,(\partial_{\tilde
  r}\,\tilde \Phi)^2+\frac{(\partial_{\tilde r}\, \ln \Theta)^2
}{8|\kappa_c|} +   \frac{5\beta}{f^4} (\partial_{\tilde r} \,\tilde
\Phi )^6\ , \\ 
 \hspace{-2mm} &  \Theta\,\partial_{\tilde r}\,\tilde \Phi
 +\frac{6\,\beta}{f^4}  \left(\partial_{\tilde r}\,\tilde \Phi
 \right)^5 = -\frac{\sqrt{|\bar \kappa_c|}}{2\pi^2}\frac{f}{\tilde r^3}\ ,  \\ 
\hspace{-2mm}&   \frac{f}{\tilde r^3} \partial_{\tilde r}
\left(\frac{\tilde r ^3}{f} \partial_{\tilde r}
  \ln\Theta\right)=2|\kappa_c| \Theta (\partial_{\tilde r } \tilde
\Phi)^2 \ , 
\end{aligned}
\ee
where we have again neglected the contributions of the Einstein-frame
potential and its derivative and defined the dimensionless variables
$\tilde r = r \, M_P$ and $\tilde \Phi= \Phi/M_P$.

As in the single-field case, $\Phi$ becomes singular inside the
instanton core $\tilde r \lesssim \tilde r_0\equiv
(|\kappa_c|^{5}\bar\kappa_c|^{-3}|\beta|)^{1/12}$ in the absence of
the higher-dimensional operator ($\beta=0$). On the other hand, we can
approximate $\Theta \sim \sigma$ there. Finally, $ \sqrt{|\bar
\kappa_c|}\tilde\Phi\sim -\sqrt{|\kappa_c|} \log(|\kappa_c|^5
|\bar\kappa_c|^{-3}\beta) \ .$

\vspace{1cm}
\setlength{\bibsep}{0.0pt}
\bibliography{SI_NP.bib}

\begin{thebibliography}{72}%
\makeatletter
\providecommand \@ifxundefined [1]{%
 \@ifx{#1\undefined}
}%
\providecommand \@ifnum [1]{%
 \ifnum #1\expandafter \@firstoftwo
 \else \expandafter \@secondoftwo
 \fi
}%
\providecommand \@ifx [1]{%
 \ifx #1\expandafter \@firstoftwo
 \else \expandafter \@secondoftwo
 \fi
}%
\providecommand \natexlab [1]{#1}%
\providecommand \enquote  [1]{``#1''}%
\providecommand \bibnamefont  [1]{#1}%
\providecommand \bibfnamefont [1]{#1}%
\providecommand \citenamefont [1]{#1}%
\providecommand \href@noop [0]{\@secondoftwo}%
\providecommand \href [0]{\begingroup \@sanitize@url \@href}%
\providecommand \@href[1]{\@@startlink{#1}\@@href}%
\providecommand \@@href[1]{\endgroup#1\@@endlink}%
\providecommand \@sanitize@url [0]{\catcode `\\12\catcode `\$12\catcode
  `\&12\catcode `\#12\catcode `\^12\catcode `\_12\catcode `\%12\relax}%
\providecommand \@@startlink[1]{}%
\providecommand \@@endlink[0]{}%
\providecommand \url  [0]{\begingroup\@sanitize@url \@url }%
\providecommand \@url [1]{\endgroup\@href {#1}{\urlprefix }}%
\providecommand \urlprefix  [0]{URL }%
\providecommand \Eprint [0]{\href }%
\providecommand \doibase [0]{http://dx.doi.org/}%
\providecommand \selectlanguage [0]{\@gobble}%
\providecommand \bibinfo  [0]{\@secondoftwo}%
\providecommand \bibfield  [0]{\@secondoftwo}%
\providecommand \translation [1]{[#1]}%
\providecommand \BibitemOpen [0]{}%
\providecommand \bibitemStop [0]{}%
\providecommand \bibitemNoStop [0]{.\EOS\space}%
\providecommand \EOS [0]{\spacefactor3000\relax}%
\providecommand \BibitemShut  [1]{\csname bibitem#1\endcsname}%
\let\auto@bib@innerbib\@empty
\bibitem [{\citenamefont {Senjanovi\'c}(2020)}]{Senjanovic:2020pnq}%
  \BibitemOpen
  \bibfield  {author} {\bibinfo {author} {\bibfnamefont {G.}~\bibnamefont
  {Senjanovi\'c}},\ }\href {\doibase 10.1142/S0217732320300062} {\bibfield
  {journal} {\bibinfo  {journal} {Mod. Phys. Lett. A}\ }\textbf {\bibinfo
  {volume} {35}},\ \bibinfo {pages} {2030006} (\bibinfo {year} {2020})},\
  \Eprint {http://arxiv.org/abs/2001.10988} {arXiv:2001.10988 [hep-ph]}
  \BibitemShut {NoStop}%
\bibitem [{\citenamefont {Feng}(2013)}]{Feng:2013pwa}%
  \BibitemOpen
  \bibfield  {author} {\bibinfo {author} {\bibfnamefont {J.~L.}\ \bibnamefont
  {Feng}},\ }\href {\doibase 10.1146/annurev-nucl-102010-130447} {\bibfield
  {journal} {\bibinfo  {journal} {Ann. Rev. Nucl. Part. Sci.}\ }\textbf
  {\bibinfo {volume} {63}},\ \bibinfo {pages} {351} (\bibinfo {year} {2013})},\
  \Eprint {http://arxiv.org/abs/1302.6587} {arXiv:1302.6587 [hep-ph]}
  \BibitemShut {NoStop}%
\bibitem [{\citenamefont {Bellazzini}\ \emph {et~al.}(2014)\citenamefont
  {Bellazzini}, \citenamefont {Csáki},\ and\ \citenamefont
  {Serra}}]{Bellazzini:2014yua}%
  \BibitemOpen
  \bibfield  {author} {\bibinfo {author} {\bibfnamefont {B.}~\bibnamefont
  {Bellazzini}}, \bibinfo {author} {\bibfnamefont {C.}~\bibnamefont {Csáki}},
  \ and\ \bibinfo {author} {\bibfnamefont {J.}~\bibnamefont {Serra}},\ }\href
  {\doibase 10.1140/epjc/s10052-014-2766-x} {\bibfield  {journal} {\bibinfo
  {journal} {Eur. Phys. J. C}\ }\textbf {\bibinfo {volume} {74}},\ \bibinfo
  {pages} {2766} (\bibinfo {year} {2014})},\ \Eprint
  {http://arxiv.org/abs/1401.2457} {arXiv:1401.2457 [hep-ph]} \BibitemShut
  {NoStop}%
\bibitem [{\citenamefont {Arkani-Hamed}\ \emph {et~al.}(1998)\citenamefont
  {Arkani-Hamed}, \citenamefont {Dimopoulos},\ and\ \citenamefont
  {Dvali}}]{ArkaniHamed:1998rs}%
  \BibitemOpen
  \bibfield  {author} {\bibinfo {author} {\bibfnamefont {N.}~\bibnamefont
  {Arkani-Hamed}}, \bibinfo {author} {\bibfnamefont {S.}~\bibnamefont
  {Dimopoulos}}, \ and\ \bibinfo {author} {\bibfnamefont {G.}~\bibnamefont
  {Dvali}},\ }\href {\doibase 10.1016/S0370-2693(98)00466-3} {\bibfield
  {journal} {\bibinfo  {journal} {Phys. Lett. B}\ }\textbf {\bibinfo {volume}
  {429}},\ \bibinfo {pages} {263} (\bibinfo {year} {1998})},\ \Eprint
  {http://arxiv.org/abs/hep-ph/9803315} {arXiv:hep-ph/9803315} \BibitemShut
  {NoStop}%
\bibitem [{\citenamefont {Dvali}\ and\ \citenamefont
  {Vilenkin}(2004)}]{Dvali:2003br}%
  \BibitemOpen
  \bibfield  {author} {\bibinfo {author} {\bibfnamefont {G.}~\bibnamefont
  {Dvali}}\ and\ \bibinfo {author} {\bibfnamefont {A.}~\bibnamefont
  {Vilenkin}},\ }\href {\doibase 10.1103/PhysRevD.70.063501} {\bibfield
  {journal} {\bibinfo  {journal} {Phys. Rev. D}\ }\textbf {\bibinfo {volume}
  {70}},\ \bibinfo {pages} {063501} (\bibinfo {year} {2004})},\ \Eprint
  {http://arxiv.org/abs/hep-th/0304043} {arXiv:hep-th/0304043} \BibitemShut
  {NoStop}%
\bibitem [{\citenamefont {Dvali}(2006)}]{Dvali:2004tma}%
  \BibitemOpen
  \bibfield  {author} {\bibinfo {author} {\bibfnamefont {G.}~\bibnamefont
  {Dvali}},\ }\href {\doibase 10.1103/PhysRevD.74.025018} {\bibfield  {journal}
  {\bibinfo  {journal} {Phys. Rev. D}\ }\textbf {\bibinfo {volume} {74}},\
  \bibinfo {pages} {025018} (\bibinfo {year} {2006})},\ \Eprint
  {http://arxiv.org/abs/hep-th/0410286} {arXiv:hep-th/0410286} \BibitemShut
  {NoStop}%
\bibitem [{\citenamefont {Dvali}(2019)}]{Dvali:2019mhn}%
  \BibitemOpen
  \bibfield  {author} {\bibinfo {author} {\bibfnamefont {G.}~\bibnamefont
  {Dvali}},\ }\href@noop {} {\  (\bibinfo {year} {2019})},\ \Eprint
  {http://arxiv.org/abs/1908.05984} {arXiv:1908.05984 [hep-ph]} \BibitemShut
  {NoStop}%
\bibitem [{\citenamefont {Michel}(2020)}]{Michel:2019vuv}%
  \BibitemOpen
  \bibfield  {author} {\bibinfo {author} {\bibfnamefont {M.}~\bibnamefont
  {Michel}},\ }\href {\doibase 10.1103/PhysRevD.101.115007} {\bibfield
  {journal} {\bibinfo  {journal} {Phys. Rev. D}\ }\textbf {\bibinfo {volume}
  {101}},\ \bibinfo {pages} {115007} (\bibinfo {year} {2020})},\ \Eprint
  {http://arxiv.org/abs/1910.10940} {arXiv:1910.10940 [hep-ph]} \BibitemShut
  {NoStop}%
\bibitem [{\citenamefont {Graham}\ \emph {et~al.}(2015)\citenamefont {Graham},
  \citenamefont {Kaplan},\ and\ \citenamefont {Rajendran}}]{Graham:2015cka}%
  \BibitemOpen
  \bibfield  {author} {\bibinfo {author} {\bibfnamefont {P.~W.}\ \bibnamefont
  {Graham}}, \bibinfo {author} {\bibfnamefont {D.~E.}\ \bibnamefont {Kaplan}},
  \ and\ \bibinfo {author} {\bibfnamefont {S.}~\bibnamefont {Rajendran}},\
  }\href {\doibase 10.1103/PhysRevLett.115.221801} {\bibfield  {journal}
  {\bibinfo  {journal} {Phys. Rev. Lett.}\ }\textbf {\bibinfo {volume} {115}},\
  \bibinfo {pages} {221801} (\bibinfo {year} {2015})},\ \Eprint
  {http://arxiv.org/abs/1504.07551} {arXiv:1504.07551 [hep-ph]} \BibitemShut
  {NoStop}%
\bibitem [{\citenamefont {Kaplan}\ and\ \citenamefont
  {Rattazzi}(2016)}]{Kaplan:2015fuy}%
  \BibitemOpen
  \bibfield  {author} {\bibinfo {author} {\bibfnamefont {D.~E.}\ \bibnamefont
  {Kaplan}}\ and\ \bibinfo {author} {\bibfnamefont {R.}~\bibnamefont
  {Rattazzi}},\ }\href {\doibase 10.1103/PhysRevD.93.085007} {\bibfield
  {journal} {\bibinfo  {journal} {Phys. Rev. D}\ }\textbf {\bibinfo {volume}
  {93}},\ \bibinfo {pages} {085007} (\bibinfo {year} {2016})},\ \Eprint
  {http://arxiv.org/abs/1511.01827} {arXiv:1511.01827 [hep-ph]} \BibitemShut
  {NoStop}%
\bibitem [{\citenamefont {Vissani}(1998)}]{Vissani:1997ys}%
  \BibitemOpen
  \bibfield  {author} {\bibinfo {author} {\bibfnamefont {F.}~\bibnamefont
  {Vissani}},\ }\href {\doibase 10.1103/PhysRevD.57.7027} {\bibfield  {journal}
  {\bibinfo  {journal} {Phys. Rev. D}\ }\textbf {\bibinfo {volume} {57}},\
  \bibinfo {pages} {7027} (\bibinfo {year} {1998})},\ \Eprint
  {http://arxiv.org/abs/hep-ph/9709409} {arXiv:hep-ph/9709409} \BibitemShut
  {NoStop}%
\bibitem [{\citenamefont {Shaposhnikov}(2007)}]{Shaposhnikov:2007nj}%
  \BibitemOpen
  \bibfield  {author} {\bibinfo {author} {\bibfnamefont {M.}~\bibnamefont
  {Shaposhnikov}},\ }in\ \href@noop {} {\emph {\bibinfo {booktitle}
  {{Astroparticle Physics: Current Issues, 2007 (APCI07)}}}}\ (\bibinfo {year}
  {2007})\ \Eprint {http://arxiv.org/abs/0708.3550} {arXiv:0708.3550 [hep-th]}
  \BibitemShut {NoStop}%
\bibitem [{\citenamefont {Shaposhnikov}\ and\ \citenamefont
  {Zenhausern}(2009{\natexlab{a}})}]{Shaposhnikov:2008xi}%
  \BibitemOpen
  \bibfield  {author} {\bibinfo {author} {\bibfnamefont {M.}~\bibnamefont
  {Shaposhnikov}}\ and\ \bibinfo {author} {\bibfnamefont {D.}~\bibnamefont
  {Zenhausern}},\ }\href {\doibase 10.1016/j.physletb.2008.11.041} {\bibfield
  {journal} {\bibinfo  {journal} {Phys. Lett. B}\ }\textbf {\bibinfo {volume}
  {671}},\ \bibinfo {pages} {162} (\bibinfo {year} {2009}{\natexlab{a}})},\
  \Eprint {http://arxiv.org/abs/0809.3406} {arXiv:0809.3406 [hep-th]}
  \BibitemShut {NoStop}%
\bibitem [{\citenamefont {Farina}\ \emph {et~al.}(2013)\citenamefont {Farina},
  \citenamefont {Pappadopulo},\ and\ \citenamefont {Strumia}}]{Farina:2013mla}%
  \BibitemOpen
  \bibfield  {author} {\bibinfo {author} {\bibfnamefont {M.}~\bibnamefont
  {Farina}}, \bibinfo {author} {\bibfnamefont {D.}~\bibnamefont {Pappadopulo}},
  \ and\ \bibinfo {author} {\bibfnamefont {A.}~\bibnamefont {Strumia}},\ }\href
  {\doibase 10.1007/JHEP08(2013)022} {\bibfield  {journal} {\bibinfo  {journal}
  {JHEP}\ }\textbf {\bibinfo {volume} {08}},\ \bibinfo {pages} {022} (\bibinfo
  {year} {2013})},\ \Eprint {http://arxiv.org/abs/1303.7244} {arXiv:1303.7244
  [hep-ph]} \BibitemShut {NoStop}%
\bibitem [{\citenamefont {Karananas}\ and\ \citenamefont
  {Shaposhnikov}(2017)}]{Karananas:2017mxm}%
  \BibitemOpen
  \bibfield  {author} {\bibinfo {author} {\bibfnamefont {G.~K.}\ \bibnamefont
  {Karananas}}\ and\ \bibinfo {author} {\bibfnamefont {M.}~\bibnamefont
  {Shaposhnikov}},\ }\href {\doibase 10.1016/j.physletb.2017.05.065} {\bibfield
   {journal} {\bibinfo  {journal} {Phys. Lett. B}\ }\textbf {\bibinfo {volume}
  {771}},\ \bibinfo {pages} {332} (\bibinfo {year} {2017})},\ \Eprint
  {http://arxiv.org/abs/1703.02964} {arXiv:1703.02964 [hep-ph]} \BibitemShut
  {NoStop}%
\bibitem [{\citenamefont {Shaposhnikov}\ and\ \citenamefont
  {Shimada}(2019)}]{Shaposhnikov:2018nnm}%
  \BibitemOpen
  \bibfield  {author} {\bibinfo {author} {\bibfnamefont {M.}~\bibnamefont
  {Shaposhnikov}}\ and\ \bibinfo {author} {\bibfnamefont {K.}~\bibnamefont
  {Shimada}},\ }\href {\doibase 10.1103/PhysRevD.99.103528} {\bibfield
  {journal} {\bibinfo  {journal} {Phys. Rev. D}\ }\textbf {\bibinfo {volume}
  {99}},\ \bibinfo {pages} {103528} (\bibinfo {year} {2019})},\ \Eprint
  {http://arxiv.org/abs/1812.08706} {arXiv:1812.08706 [hep-ph]} \BibitemShut
  {NoStop}%
\bibitem [{\citenamefont {Wetterich}(2012)}]{Wetterich:2011aa}%
  \BibitemOpen
  \bibfield  {author} {\bibinfo {author} {\bibfnamefont {C.}~\bibnamefont
  {Wetterich}},\ }\href {\doibase 10.1016/j.physletb.2012.11.020} {\bibfield
  {journal} {\bibinfo  {journal} {Phys. Lett. B}\ }\textbf {\bibinfo {volume}
  {718}},\ \bibinfo {pages} {573} (\bibinfo {year} {2012})},\ \Eprint
  {http://arxiv.org/abs/1112.2910} {arXiv:1112.2910 [hep-ph]} \BibitemShut
  {NoStop}%
\bibitem [{\citenamefont {Wetterich}\ and\ \citenamefont
  {Yamada}(2017)}]{Wetterich:2016uxm}%
  \BibitemOpen
  \bibfield  {author} {\bibinfo {author} {\bibfnamefont {C.}~\bibnamefont
  {Wetterich}}\ and\ \bibinfo {author} {\bibfnamefont {M.}~\bibnamefont
  {Yamada}},\ }\href {\doibase 10.1016/j.physletb.2017.04.049} {\bibfield
  {journal} {\bibinfo  {journal} {Phys. Lett. B}\ }\textbf {\bibinfo {volume}
  {770}},\ \bibinfo {pages} {268} (\bibinfo {year} {2017})},\ \Eprint
  {http://arxiv.org/abs/1612.03069} {arXiv:1612.03069 [hep-th]} \BibitemShut
  {NoStop}%
\bibitem [{\citenamefont {Dvali}\ and\ \citenamefont
  {Gomez}(2010)}]{Dvali:2010bf}%
  \BibitemOpen
  \bibfield  {author} {\bibinfo {author} {\bibfnamefont {G.}~\bibnamefont
  {Dvali}}\ and\ \bibinfo {author} {\bibfnamefont {C.}~\bibnamefont {Gomez}},\
  }\href@noop {} {\  (\bibinfo {year} {2010})},\ \Eprint
  {http://arxiv.org/abs/1005.3497} {arXiv:1005.3497 [hep-th]} \BibitemShut
  {NoStop}%
\bibitem [{\citenamefont {Dvali}\ \emph {et~al.}(2011)\citenamefont {Dvali},
  \citenamefont {Folkerts},\ and\ \citenamefont {Germani}}]{Dvali:2010ue}%
  \BibitemOpen
  \bibfield  {author} {\bibinfo {author} {\bibfnamefont {G.}~\bibnamefont
  {Dvali}}, \bibinfo {author} {\bibfnamefont {S.}~\bibnamefont {Folkerts}}, \
  and\ \bibinfo {author} {\bibfnamefont {C.}~\bibnamefont {Germani}},\ }\href
  {\doibase 10.1103/PhysRevD.84.024039} {\bibfield  {journal} {\bibinfo
  {journal} {Phys. Rev. D}\ }\textbf {\bibinfo {volume} {84}},\ \bibinfo
  {pages} {024039} (\bibinfo {year} {2011})},\ \Eprint
  {http://arxiv.org/abs/1006.0984} {arXiv:1006.0984 [hep-th]} \BibitemShut
  {NoStop}%
\bibitem [{\citenamefont {Shaposhnikov}\ and\ \citenamefont
  {Shkerin}(2018{\natexlab{a}})}]{Shaposhnikov:2018xkv}%
  \BibitemOpen
  \bibfield  {author} {\bibinfo {author} {\bibfnamefont {M.}~\bibnamefont
  {Shaposhnikov}}\ and\ \bibinfo {author} {\bibfnamefont {A.}~\bibnamefont
  {Shkerin}},\ }\href {\doibase 10.1016/j.physletb.2018.06.068} {\bibfield
  {journal} {\bibinfo  {journal} {Phys. Lett.}\ }\textbf {\bibinfo {volume}
  {B783}},\ \bibinfo {pages} {253} (\bibinfo {year} {2018}{\natexlab{a}})},\
  \Eprint {http://arxiv.org/abs/1803.08907} {arXiv:1803.08907 [hep-th]}
  \BibitemShut {NoStop}%
\bibitem [{\citenamefont {Shaposhnikov}\ and\ \citenamefont
  {Shkerin}(2018{\natexlab{b}})}]{Shaposhnikov:2018jag}%
  \BibitemOpen
  \bibfield  {author} {\bibinfo {author} {\bibfnamefont {M.}~\bibnamefont
  {Shaposhnikov}}\ and\ \bibinfo {author} {\bibfnamefont {A.}~\bibnamefont
  {Shkerin}},\ }\href {\doibase 10.1007/JHEP10(2018)024} {\bibfield  {journal}
  {\bibinfo  {journal} {JHEP}\ }\textbf {\bibinfo {volume} {10}},\ \bibinfo
  {pages} {024} (\bibinfo {year} {2018}{\natexlab{b}})},\ \Eprint
  {http://arxiv.org/abs/1804.06376} {arXiv:1804.06376 [hep-th]} \BibitemShut
  {NoStop}%
\bibitem [{\citenamefont {Shkerin}(2019)}]{Shkerin:2019mmu}%
  \BibitemOpen
  \bibfield  {author} {\bibinfo {author} {\bibfnamefont {A.}~\bibnamefont
  {Shkerin}},\ }\href {\doibase 10.1103/PhysRevD.99.115018} {\bibfield
  {journal} {\bibinfo  {journal} {Phys. Rev.}\ }\textbf {\bibinfo {volume}
  {D99}},\ \bibinfo {pages} {115018} (\bibinfo {year} {2019})},\ \Eprint
  {http://arxiv.org/abs/1903.11317} {arXiv:1903.11317 [hep-th]} \BibitemShut
  {NoStop}%
\bibitem [{\citenamefont {Shaposhnikov}\ \emph
  {et~al.}(2020{\natexlab{a}})\citenamefont {Shaposhnikov}, \citenamefont
  {Shkerin},\ and\ \citenamefont {Zell}}]{Shaposhnikov:2020geh}%
  \BibitemOpen
  \bibfield  {author} {\bibinfo {author} {\bibfnamefont {M.}~\bibnamefont
  {Shaposhnikov}}, \bibinfo {author} {\bibfnamefont {A.}~\bibnamefont
  {Shkerin}}, \ and\ \bibinfo {author} {\bibfnamefont {S.}~\bibnamefont
  {Zell}},\ }\href@noop {} {\  (\bibinfo {year} {2020}{\natexlab{a}})},\
  \Eprint {http://arxiv.org/abs/2001.09088} {arXiv:2001.09088 [hep-th]}
  \BibitemShut {NoStop}%
\bibitem [{\citenamefont {Akrami}\ \emph {et~al.}(2018)\citenamefont {Akrami}
  \emph {et~al.}}]{Akrami:2018odb}%
  \BibitemOpen
  \bibfield  {author} {\bibinfo {author} {\bibfnamefont {Y.}~\bibnamefont
  {Akrami}} \emph {et~al.} (\bibinfo {collaboration} {Planck}),\ }\href@noop {}
  {\  (\bibinfo {year} {2018})},\ \Eprint {http://arxiv.org/abs/1807.06211}
  {arXiv:1807.06211 [astro-ph.CO]} \BibitemShut {NoStop}%
\bibitem [{\citenamefont {Shaposhnikov}\ \emph
  {et~al.}(2020{\natexlab{b}})\citenamefont {Shaposhnikov}, \citenamefont
  {Shkerin},\ and\ \citenamefont {Zell}}]{Shaposhnikov:2020fdv}%
  \BibitemOpen
  \bibfield  {author} {\bibinfo {author} {\bibfnamefont {M.}~\bibnamefont
  {Shaposhnikov}}, \bibinfo {author} {\bibfnamefont {A.}~\bibnamefont
  {Shkerin}}, \ and\ \bibinfo {author} {\bibfnamefont {S.}~\bibnamefont
  {Zell}},\ }\href {\doibase 10.1088/1475-7516/2020/07/064} {\bibfield
  {journal} {\bibinfo  {journal} {JCAP}\ }\textbf {\bibinfo {volume} {07}},\
  \bibinfo {pages} {064} (\bibinfo {year} {2020}{\natexlab{b}})},\ \Eprint
  {http://arxiv.org/abs/2002.07105} {arXiv:2002.07105 [hep-ph]} \BibitemShut
  {NoStop}%
\bibitem [{\citenamefont {Kallosh}\ and\ \citenamefont
  {Linde}(2013)}]{Kallosh:2013hoa}%
  \BibitemOpen
  \bibfield  {author} {\bibinfo {author} {\bibfnamefont {R.}~\bibnamefont
  {Kallosh}}\ and\ \bibinfo {author} {\bibfnamefont {A.}~\bibnamefont
  {Linde}},\ }\href {\doibase 10.1088/1475-7516/2013/07/002} {\bibfield
  {journal} {\bibinfo  {journal} {JCAP}\ }\textbf {\bibinfo {volume} {07}},\
  \bibinfo {pages} {002} (\bibinfo {year} {2013})},\ \Eprint
  {http://arxiv.org/abs/1306.5220} {arXiv:1306.5220 [hep-th]} \BibitemShut
  {NoStop}%
\bibitem [{\citenamefont {Galante}\ \emph {et~al.}(2015)\citenamefont
  {Galante}, \citenamefont {Kallosh}, \citenamefont {Linde},\ and\
  \citenamefont {Roest}}]{Galante:2014ifa}%
  \BibitemOpen
  \bibfield  {author} {\bibinfo {author} {\bibfnamefont {M.}~\bibnamefont
  {Galante}}, \bibinfo {author} {\bibfnamefont {R.}~\bibnamefont {Kallosh}},
  \bibinfo {author} {\bibfnamefont {A.}~\bibnamefont {Linde}}, \ and\ \bibinfo
  {author} {\bibfnamefont {D.}~\bibnamefont {Roest}},\ }\href {\doibase
  10.1103/PhysRevLett.114.141302} {\bibfield  {journal} {\bibinfo  {journal}
  {Phys. Rev. Lett.}\ }\textbf {\bibinfo {volume} {114}},\ \bibinfo {pages}
  {141302} (\bibinfo {year} {2015})},\ \Eprint {http://arxiv.org/abs/1412.3797}
  {arXiv:1412.3797 [hep-th]} \BibitemShut {NoStop}%
\bibitem [{\citenamefont {Kallosh}\ and\ \citenamefont
  {Linde}(2015)}]{Kallosh:2015zsa}%
  \BibitemOpen
  \bibfield  {author} {\bibinfo {author} {\bibfnamefont {R.}~\bibnamefont
  {Kallosh}}\ and\ \bibinfo {author} {\bibfnamefont {A.}~\bibnamefont
  {Linde}},\ }\href {\doibase 10.1016/j.crhy.2015.07.004} {\bibfield  {journal}
  {\bibinfo  {journal} {Comptes Rendus Physique}\ }\textbf {\bibinfo {volume}
  {16}},\ \bibinfo {pages} {914} (\bibinfo {year} {2015})},\ \Eprint
  {http://arxiv.org/abs/1503.06785} {arXiv:1503.06785 [hep-th]} \BibitemShut
  {NoStop}%
\bibitem [{\citenamefont {Artymowski}\ and\ \citenamefont
  {Rubio}(2016)}]{Artymowski:2016pjz}%
  \BibitemOpen
  \bibfield  {author} {\bibinfo {author} {\bibfnamefont {M.}~\bibnamefont
  {Artymowski}}\ and\ \bibinfo {author} {\bibfnamefont {J.}~\bibnamefont
  {Rubio}},\ }\href {\doibase 10.1016/j.physletb.2016.08.024} {\bibfield
  {journal} {\bibinfo  {journal} {Phys. Lett. B}\ }\textbf {\bibinfo {volume}
  {761}},\ \bibinfo {pages} {111} (\bibinfo {year} {2016})},\ \Eprint
  {http://arxiv.org/abs/1607.00398} {arXiv:1607.00398 [astro-ph.CO]}
  \BibitemShut {NoStop}%
\bibitem [{\citenamefont {Fumagalli}(2017)}]{Fumagalli:2016sof}%
  \BibitemOpen
  \bibfield  {author} {\bibinfo {author} {\bibfnamefont {J.}~\bibnamefont
  {Fumagalli}},\ }\href {\doibase 10.1016/j.physletb.2017.04.017} {\bibfield
  {journal} {\bibinfo  {journal} {Phys. Lett. B}\ }\textbf {\bibinfo {volume}
  {769}},\ \bibinfo {pages} {451} (\bibinfo {year} {2017})},\ \Eprint
  {http://arxiv.org/abs/1611.04997} {arXiv:1611.04997 [hep-th]} \BibitemShut
  {NoStop}%
\bibitem [{\citenamefont {Cremmer}\ \emph {et~al.}(1978)\citenamefont
  {Cremmer}, \citenamefont {Scherk},\ and\ \citenamefont
  {Ferrara}}]{Cremmer:1977tt}%
  \BibitemOpen
  \bibfield  {author} {\bibinfo {author} {\bibfnamefont {E.}~\bibnamefont
  {Cremmer}}, \bibinfo {author} {\bibfnamefont {J.}~\bibnamefont {Scherk}}, \
  and\ \bibinfo {author} {\bibfnamefont {S.}~\bibnamefont {Ferrara}},\ }\href
  {\doibase 10.1016/0370-2693(78)90060-6} {\bibfield  {journal} {\bibinfo
  {journal} {Phys. Lett. B}\ }\textbf {\bibinfo {volume} {74}},\ \bibinfo
  {pages} {61} (\bibinfo {year} {1978})}\BibitemShut {NoStop}%
\bibitem [{\citenamefont {Bergshoeff}\ \emph {et~al.}(1981)\citenamefont
  {Bergshoeff}, \citenamefont {de~Roo},\ and\ \citenamefont
  {de~Wit}}]{Bergshoeff:1980is}%
  \BibitemOpen
  \bibfield  {author} {\bibinfo {author} {\bibfnamefont {E.}~\bibnamefont
  {Bergshoeff}}, \bibinfo {author} {\bibfnamefont {M.}~\bibnamefont {de~Roo}},
  \ and\ \bibinfo {author} {\bibfnamefont {B.}~\bibnamefont {de~Wit}},\ }\href
  {\doibase 10.1016/0550-3213(81)90465-X} {\bibfield  {journal} {\bibinfo
  {journal} {Nucl. Phys. B}\ }\textbf {\bibinfo {volume} {182}},\ \bibinfo
  {pages} {173} (\bibinfo {year} {1981})}\BibitemShut {NoStop}%
\bibitem [{\citenamefont {de~Roo}(1985)}]{deRoo:1984zyh}%
  \BibitemOpen
  \bibfield  {author} {\bibinfo {author} {\bibfnamefont {M.}~\bibnamefont
  {de~Roo}},\ }\href {\doibase 10.1016/0550-3213(85)90151-8} {\bibfield
  {journal} {\bibinfo  {journal} {Nucl. Phys. B}\ }\textbf {\bibinfo {volume}
  {255}},\ \bibinfo {pages} {515} (\bibinfo {year} {1985})}\BibitemShut
  {NoStop}%
\bibitem [{\citenamefont {Ferrara}\ \emph {et~al.}(2013)\citenamefont
  {Ferrara}, \citenamefont {Kallosh},\ and\ \citenamefont
  {Van~Proeyen}}]{Ferrara:2012ui}%
  \BibitemOpen
  \bibfield  {author} {\bibinfo {author} {\bibfnamefont {S.}~\bibnamefont
  {Ferrara}}, \bibinfo {author} {\bibfnamefont {R.}~\bibnamefont {Kallosh}}, \
  and\ \bibinfo {author} {\bibfnamefont {A.}~\bibnamefont {Van~Proeyen}},\
  }\href {\doibase 10.1103/PhysRevD.87.025004} {\bibfield  {journal} {\bibinfo
  {journal} {Phys. Rev. D}\ }\textbf {\bibinfo {volume} {87}},\ \bibinfo
  {pages} {025004} (\bibinfo {year} {2013})},\ \Eprint
  {http://arxiv.org/abs/1209.0418} {arXiv:1209.0418 [hep-th]} \BibitemShut
  {NoStop}%
\bibitem [{\citenamefont {'t~Hooft}(1973)}]{tHooft:1973mfk}%
  \BibitemOpen
  \bibfield  {author} {\bibinfo {author} {\bibfnamefont {G.}~\bibnamefont
  {'t~Hooft}},\ }\href {\doibase 10.1016/0550-3213(73)90376-3} {\bibfield
  {journal} {\bibinfo  {journal} {Nucl. Phys. B}\ }\textbf {\bibinfo {volume}
  {61}},\ \bibinfo {pages} {455} (\bibinfo {year} {1973})}\BibitemShut
  {NoStop}%
\bibitem [{\citenamefont {Weinberg}(1973)}]{Weinberg:1973am}%
  \BibitemOpen
  \bibfield  {author} {\bibinfo {author} {\bibfnamefont {E.~J.}\ \bibnamefont
  {Weinberg}},\ }\emph {\bibinfo {title} {{Radiative corrections as the origin
  of spontaneous symmetry breaking}}},\ \href@noop {} {Ph.D. thesis},\ \bibinfo
   {school} {Harvard U.} (\bibinfo {year} {1973}),\ \Eprint
  {http://arxiv.org/abs/hep-th/0507214} {arXiv:hep-th/0507214} \BibitemShut
  {NoStop}%
\bibitem [{\citenamefont {Weinberg}(1976)}]{Weinberg:1976pe}%
  \BibitemOpen
  \bibfield  {author} {\bibinfo {author} {\bibfnamefont {S.}~\bibnamefont
  {Weinberg}},\ }\href {\doibase 10.1103/PhysRevLett.36.294} {\bibfield
  {journal} {\bibinfo  {journal} {Phys. Rev. Lett.}\ }\textbf {\bibinfo
  {volume} {36}},\ \bibinfo {pages} {294} (\bibinfo {year} {1976})}\BibitemShut
  {NoStop}%
\bibitem [{\citenamefont {Linde}(1977)}]{Linde:1977mm}%
  \BibitemOpen
  \bibfield  {author} {\bibinfo {author} {\bibfnamefont {A.~D.}\ \bibnamefont
  {Linde}},\ }\href {\doibase 10.1016/0370-2693(77)90664-5} {\bibfield
  {journal} {\bibinfo  {journal} {Phys. Lett. B}\ }\textbf {\bibinfo {volume}
  {70}},\ \bibinfo {pages} {306} (\bibinfo {year} {1977})}\BibitemShut
  {NoStop}%
\bibitem [{\citenamefont {Rasanen}(2019)}]{Rasanen:2018ihz}%
  \BibitemOpen
  \bibfield  {author} {\bibinfo {author} {\bibfnamefont {S.}~\bibnamefont
  {Rasanen}},\ }\href {\doibase 10.21105/astro.1811.09514} {\bibfield
  {journal} {\bibinfo  {journal} {Open J. Astrophys.}\ }\textbf {\bibinfo
  {volume} {2}},\ \bibinfo {pages} {1} (\bibinfo {year} {2019})},\ \Eprint
  {http://arxiv.org/abs/1811.09514} {arXiv:1811.09514 [gr-qc]} \BibitemShut
  {NoStop}%
\bibitem [{\citenamefont {Karananas}\ and\ \citenamefont
  {Rubio}(2016)}]{Karananas:2016kyt}%
  \BibitemOpen
  \bibfield  {author} {\bibinfo {author} {\bibfnamefont {G.~K.}\ \bibnamefont
  {Karananas}}\ and\ \bibinfo {author} {\bibfnamefont {J.}~\bibnamefont
  {Rubio}},\ }\href {\doibase 10.1016/j.physletb.2016.08.037} {\bibfield
  {journal} {\bibinfo  {journal} {Phys. Lett.}\ }\textbf {\bibinfo {volume}
  {B761}},\ \bibinfo {pages} {223} (\bibinfo {year} {2016})},\ \Eprint
  {http://arxiv.org/abs/1606.08848} {arXiv:1606.08848 [hep-ph]} \BibitemShut
  {NoStop}%
\bibitem [{\citenamefont {Karananas}\ and\ \citenamefont
  {Shaposhnikov}(2016)}]{Karananas:2016grc}%
  \BibitemOpen
  \bibfield  {author} {\bibinfo {author} {\bibfnamefont {G.~K.}\ \bibnamefont
  {Karananas}}\ and\ \bibinfo {author} {\bibfnamefont {M.}~\bibnamefont
  {Shaposhnikov}},\ }\href {\doibase 10.1103/PhysRevD.93.084052} {\bibfield
  {journal} {\bibinfo  {journal} {Phys. Rev. D}\ }\textbf {\bibinfo {volume}
  {93}},\ \bibinfo {pages} {084052} (\bibinfo {year} {2016})},\ \Eprint
  {http://arxiv.org/abs/1603.01274} {arXiv:1603.01274 [hep-th]} \BibitemShut
  {NoStop}%
\bibitem [{\citenamefont {Casas}\ \emph {et~al.}(2018)\citenamefont {Casas},
  \citenamefont {Pauly},\ and\ \citenamefont {Rubio}}]{Casas:2017wjh}%
  \BibitemOpen
  \bibfield  {author} {\bibinfo {author} {\bibfnamefont {S.}~\bibnamefont
  {Casas}}, \bibinfo {author} {\bibfnamefont {M.}~\bibnamefont {Pauly}}, \ and\
  \bibinfo {author} {\bibfnamefont {J.}~\bibnamefont {Rubio}},\ }\href
  {\doibase 10.1103/PhysRevD.97.043520} {\bibfield  {journal} {\bibinfo
  {journal} {Phys. Rev.}\ }\textbf {\bibinfo {volume} {D97}},\ \bibinfo {pages}
  {043520} (\bibinfo {year} {2018})},\ \Eprint
  {http://arxiv.org/abs/1712.04956} {arXiv:1712.04956 [astro-ph.CO]}
  \BibitemShut {NoStop}%
\bibitem [{\citenamefont {Rubio}(2019)}]{Rubio:2018ogq}%
  \BibitemOpen
  \bibfield  {author} {\bibinfo {author} {\bibfnamefont {J.}~\bibnamefont
  {Rubio}},\ }\href {\doibase 10.3389/fspas.2018.00050} {\bibfield  {journal}
  {\bibinfo  {journal} {Front. Astron. Space Sci.}\ }\textbf {\bibinfo {volume}
  {5}},\ \bibinfo {pages} {50} (\bibinfo {year} {2019})},\ \Eprint
  {http://arxiv.org/abs/1807.02376} {arXiv:1807.02376 [hep-ph]} \BibitemShut
  {NoStop}%
\bibitem [{\citenamefont {Garcia-Bellido}\ \emph {et~al.}(2009)\citenamefont
  {Garcia-Bellido}, \citenamefont {Figueroa},\ and\ \citenamefont
  {Rubio}}]{GarciaBellido:2008ab}%
  \BibitemOpen
  \bibfield  {author} {\bibinfo {author} {\bibfnamefont {J.}~\bibnamefont
  {Garcia-Bellido}}, \bibinfo {author} {\bibfnamefont {D.~G.}\ \bibnamefont
  {Figueroa}}, \ and\ \bibinfo {author} {\bibfnamefont {J.}~\bibnamefont
  {Rubio}},\ }\href {\doibase 10.1103/PhysRevD.79.063531} {\bibfield  {journal}
  {\bibinfo  {journal} {Phys. Rev. D}\ }\textbf {\bibinfo {volume} {79}},\
  \bibinfo {pages} {063531} (\bibinfo {year} {2009})},\ \Eprint
  {http://arxiv.org/abs/0812.4624} {arXiv:0812.4624 [hep-ph]} \BibitemShut
  {NoStop}%
\bibitem [{\citenamefont {Bezrukov}\ \emph {et~al.}(2009)\citenamefont
  {Bezrukov}, \citenamefont {Gorbunov},\ and\ \citenamefont
  {Shaposhnikov}}]{Bezrukov:2008ut}%
  \BibitemOpen
  \bibfield  {author} {\bibinfo {author} {\bibfnamefont {F.}~\bibnamefont
  {Bezrukov}}, \bibinfo {author} {\bibfnamefont {D.}~\bibnamefont {Gorbunov}},
  \ and\ \bibinfo {author} {\bibfnamefont {M.}~\bibnamefont {Shaposhnikov}},\
  }\href {\doibase 10.1088/1475-7516/2009/06/029} {\bibfield  {journal}
  {\bibinfo  {journal} {JCAP}\ }\textbf {\bibinfo {volume} {06}},\ \bibinfo
  {pages} {029} (\bibinfo {year} {2009})},\ \Eprint
  {http://arxiv.org/abs/0812.3622} {arXiv:0812.3622 [hep-ph]} \BibitemShut
  {NoStop}%
\bibitem [{\citenamefont {Repond}\ and\ \citenamefont
  {Rubio}(2016)}]{Repond:2016sol}%
  \BibitemOpen
  \bibfield  {author} {\bibinfo {author} {\bibfnamefont {J.}~\bibnamefont
  {Repond}}\ and\ \bibinfo {author} {\bibfnamefont {J.}~\bibnamefont {Rubio}},\
  }\href {\doibase 10.1088/1475-7516/2016/07/043} {\bibfield  {journal}
  {\bibinfo  {journal} {JCAP}\ }\textbf {\bibinfo {volume} {07}},\ \bibinfo
  {pages} {043} (\bibinfo {year} {2016})},\ \Eprint
  {http://arxiv.org/abs/1604.08238} {arXiv:1604.08238 [astro-ph.CO]}
  \BibitemShut {NoStop}%
\bibitem [{\citenamefont {Rubio}\ and\ \citenamefont
  {Tomberg}(2019)}]{Rubio:2019ypq}%
  \BibitemOpen
  \bibfield  {author} {\bibinfo {author} {\bibfnamefont {J.}~\bibnamefont
  {Rubio}}\ and\ \bibinfo {author} {\bibfnamefont {E.~S.}\ \bibnamefont
  {Tomberg}},\ }\href {\doibase 10.1088/1475-7516/2019/04/021} {\bibfield
  {journal} {\bibinfo  {journal} {JCAP}\ }\textbf {\bibinfo {volume} {04}},\
  \bibinfo {pages} {021} (\bibinfo {year} {2019})},\ \Eprint
  {http://arxiv.org/abs/1902.10148} {arXiv:1902.10148 [hep-ph]} \BibitemShut
  {NoStop}%
\bibitem [{\citenamefont {Bezrukov}\ and\ \citenamefont
  {Shaposhnikov}(2008)}]{Bezrukov:2007ep}%
  \BibitemOpen
  \bibfield  {author} {\bibinfo {author} {\bibfnamefont {F.~L.}\ \bibnamefont
  {Bezrukov}}\ and\ \bibinfo {author} {\bibfnamefont {M.}~\bibnamefont
  {Shaposhnikov}},\ }\href {\doibase 10.1016/j.physletb.2007.11.072} {\bibfield
   {journal} {\bibinfo  {journal} {Phys. Lett. B}\ }\textbf {\bibinfo {volume}
  {659}},\ \bibinfo {pages} {703} (\bibinfo {year} {2008})},\ \Eprint
  {http://arxiv.org/abs/0710.3755} {arXiv:0710.3755 [hep-th]} \BibitemShut
  {NoStop}%
\bibitem [{\citenamefont {Bauer}\ and\ \citenamefont
  {Demir}(2008)}]{Bauer:2008zj}%
  \BibitemOpen
  \bibfield  {author} {\bibinfo {author} {\bibfnamefont {F.}~\bibnamefont
  {Bauer}}\ and\ \bibinfo {author} {\bibfnamefont {D.~A.}\ \bibnamefont
  {Demir}},\ }\href {\doibase 10.1016/j.physletb.2008.06.014} {\bibfield
  {journal} {\bibinfo  {journal} {Phys. Lett. B}\ }\textbf {\bibinfo {volume}
  {665}},\ \bibinfo {pages} {222} (\bibinfo {year} {2008})},\ \Eprint
  {http://arxiv.org/abs/0803.2664} {arXiv:0803.2664 [hep-ph]} \BibitemShut
  {NoStop}%
\bibitem [{\citenamefont {Bezrukov}\ \emph {et~al.}(2018)\citenamefont
  {Bezrukov}, \citenamefont {Pauly},\ and\ \citenamefont
  {Rubio}}]{Bezrukov:2017dyv}%
  \BibitemOpen
  \bibfield  {author} {\bibinfo {author} {\bibfnamefont {F.}~\bibnamefont
  {Bezrukov}}, \bibinfo {author} {\bibfnamefont {M.}~\bibnamefont {Pauly}}, \
  and\ \bibinfo {author} {\bibfnamefont {J.}~\bibnamefont {Rubio}},\ }\href
  {\doibase 10.1088/1475-7516/2018/02/040} {\bibfield  {journal} {\bibinfo
  {journal} {JCAP}\ }\textbf {\bibinfo {volume} {02}},\ \bibinfo {pages} {040}
  (\bibinfo {year} {2018})},\ \Eprint {http://arxiv.org/abs/1706.05007}
  {arXiv:1706.05007 [hep-ph]} \BibitemShut {NoStop}%
\bibitem [{\citenamefont {Fumagalli}\ and\ \citenamefont
  {Postma}(2016)}]{Fumagalli:2016lls}%
  \BibitemOpen
  \bibfield  {author} {\bibinfo {author} {\bibfnamefont {J.}~\bibnamefont
  {Fumagalli}}\ and\ \bibinfo {author} {\bibfnamefont {M.}~\bibnamefont
  {Postma}},\ }\href {\doibase 10.1007/JHEP05(2016)049} {\bibfield  {journal}
  {\bibinfo  {journal} {JHEP}\ }\textbf {\bibinfo {volume} {05}},\ \bibinfo
  {pages} {049} (\bibinfo {year} {2016})},\ \Eprint
  {http://arxiv.org/abs/1602.07234} {arXiv:1602.07234 [hep-ph]} \BibitemShut
  {NoStop}%
\bibitem [{\citenamefont {Rasanen}\ and\ \citenamefont
  {Wahlman}(2017)}]{Rasanen:2017ivk}%
  \BibitemOpen
  \bibfield  {author} {\bibinfo {author} {\bibfnamefont {S.}~\bibnamefont
  {Rasanen}}\ and\ \bibinfo {author} {\bibfnamefont {P.}~\bibnamefont
  {Wahlman}},\ }\href {\doibase 10.1088/1475-7516/2017/11/047} {\bibfield
  {journal} {\bibinfo  {journal} {JCAP}\ }\textbf {\bibinfo {volume} {11}},\
  \bibinfo {pages} {047} (\bibinfo {year} {2017})},\ \Eprint
  {http://arxiv.org/abs/1709.07853} {arXiv:1709.07853 [astro-ph.CO]}
  \BibitemShut {NoStop}%
\bibitem [{\citenamefont {Bezrukov}\ \emph {et~al.}(2011)\citenamefont
  {Bezrukov}, \citenamefont {Magnin}, \citenamefont {Shaposhnikov},\ and\
  \citenamefont {Sibiryakov}}]{Bezrukov:2010jz}%
  \BibitemOpen
  \bibfield  {author} {\bibinfo {author} {\bibfnamefont {F.}~\bibnamefont
  {Bezrukov}}, \bibinfo {author} {\bibfnamefont {A.}~\bibnamefont {Magnin}},
  \bibinfo {author} {\bibfnamefont {M.}~\bibnamefont {Shaposhnikov}}, \ and\
  \bibinfo {author} {\bibfnamefont {S.}~\bibnamefont {Sibiryakov}},\ }\href
  {\doibase 10.1007/JHEP01(2011)016} {\bibfield  {journal} {\bibinfo  {journal}
  {JHEP}\ }\textbf {\bibinfo {volume} {01}},\ \bibinfo {pages} {016} (\bibinfo
  {year} {2011})},\ \Eprint {http://arxiv.org/abs/1008.5157} {arXiv:1008.5157
  [hep-ph]} \BibitemShut {NoStop}%
\bibitem [{\citenamefont {Bezrukov}\ \emph {et~al.}(2015)\citenamefont
  {Bezrukov}, \citenamefont {Rubio},\ and\ \citenamefont
  {Shaposhnikov}}]{Bezrukov:2014ipa}%
  \BibitemOpen
  \bibfield  {author} {\bibinfo {author} {\bibfnamefont {F.}~\bibnamefont
  {Bezrukov}}, \bibinfo {author} {\bibfnamefont {J.}~\bibnamefont {Rubio}}, \
  and\ \bibinfo {author} {\bibfnamefont {M.}~\bibnamefont {Shaposhnikov}},\
  }\href {\doibase 10.1103/PhysRevD.92.083512} {\bibfield  {journal} {\bibinfo
  {journal} {Phys. Rev. D}\ }\textbf {\bibinfo {volume} {92}},\ \bibinfo
  {pages} {083512} (\bibinfo {year} {2015})},\ \Eprint
  {http://arxiv.org/abs/1412.3811} {arXiv:1412.3811 [hep-ph]} \BibitemShut
  {NoStop}%
\bibitem [{\citenamefont {Shaposhnikov}\ and\ \citenamefont
  {Zenhausern}(2009{\natexlab{b}})}]{Shaposhnikov:2008xb}%
  \BibitemOpen
  \bibfield  {author} {\bibinfo {author} {\bibfnamefont {M.}~\bibnamefont
  {Shaposhnikov}}\ and\ \bibinfo {author} {\bibfnamefont {D.}~\bibnamefont
  {Zenhausern}},\ }\href {\doibase 10.1016/j.physletb.2008.11.054} {\bibfield
  {journal} {\bibinfo  {journal} {Phys. Lett. B}\ }\textbf {\bibinfo {volume}
  {671}},\ \bibinfo {pages} {187} (\bibinfo {year} {2009}{\natexlab{b}})},\
  \Eprint {http://arxiv.org/abs/0809.3395} {arXiv:0809.3395 [hep-th]}
  \BibitemShut {NoStop}%
\bibitem [{\citenamefont {Garcia-Bellido}\ \emph {et~al.}(2011)\citenamefont
  {Garcia-Bellido}, \citenamefont {Rubio}, \citenamefont {Shaposhnikov},\ and\
  \citenamefont {Zenhausern}}]{GarciaBellido:2011de}%
  \BibitemOpen
  \bibfield  {author} {\bibinfo {author} {\bibfnamefont {J.}~\bibnamefont
  {Garcia-Bellido}}, \bibinfo {author} {\bibfnamefont {J.}~\bibnamefont
  {Rubio}}, \bibinfo {author} {\bibfnamefont {M.}~\bibnamefont {Shaposhnikov}},
  \ and\ \bibinfo {author} {\bibfnamefont {D.}~\bibnamefont {Zenhausern}},\
  }\href {\doibase 10.1103/PhysRevD.84.123504} {\bibfield  {journal} {\bibinfo
  {journal} {Phys. Rev. D}\ }\textbf {\bibinfo {volume} {84}},\ \bibinfo
  {pages} {123504} (\bibinfo {year} {2011})},\ \Eprint
  {http://arxiv.org/abs/1107.2163} {arXiv:1107.2163 [hep-ph]} \BibitemShut
  {NoStop}%
\bibitem [{\citenamefont {Bezrukov}\ \emph {et~al.}(2013)\citenamefont
  {Bezrukov}, \citenamefont {Karananas}, \citenamefont {Rubio},\ and\
  \citenamefont {Shaposhnikov}}]{Bezrukov:2012hx}%
  \BibitemOpen
  \bibfield  {author} {\bibinfo {author} {\bibfnamefont {F.}~\bibnamefont
  {Bezrukov}}, \bibinfo {author} {\bibfnamefont {G.~K.}\ \bibnamefont
  {Karananas}}, \bibinfo {author} {\bibfnamefont {J.}~\bibnamefont {Rubio}}, \
  and\ \bibinfo {author} {\bibfnamefont {M.}~\bibnamefont {Shaposhnikov}},\
  }\href {\doibase 10.1103/PhysRevD.87.096001} {\bibfield  {journal} {\bibinfo
  {journal} {Phys. Rev. D}\ }\textbf {\bibinfo {volume} {87}},\ \bibinfo
  {pages} {096001} (\bibinfo {year} {2013})},\ \Eprint
  {http://arxiv.org/abs/1212.4148} {arXiv:1212.4148 [hep-ph]} \BibitemShut
  {NoStop}%
\bibitem [{\citenamefont {Garcia-Bellido}\ \emph {et~al.}(2012)\citenamefont
  {Garcia-Bellido}, \citenamefont {Rubio},\ and\ \citenamefont
  {Shaposhnikov}}]{GarciaBellido:2012zu}%
  \BibitemOpen
  \bibfield  {author} {\bibinfo {author} {\bibfnamefont {J.}~\bibnamefont
  {Garcia-Bellido}}, \bibinfo {author} {\bibfnamefont {J.}~\bibnamefont
  {Rubio}}, \ and\ \bibinfo {author} {\bibfnamefont {M.}~\bibnamefont
  {Shaposhnikov}},\ }\href {\doibase 10.1016/j.physletb.2012.10.075} {\bibfield
   {journal} {\bibinfo  {journal} {Phys. Lett. B}\ }\textbf {\bibinfo {volume}
  {718}},\ \bibinfo {pages} {507} (\bibinfo {year} {2012})},\ \Eprint
  {http://arxiv.org/abs/1209.2119} {arXiv:1209.2119 [hep-ph]} \BibitemShut
  {NoStop}%
\bibitem [{\citenamefont {Rubio}\ and\ \citenamefont
  {Shaposhnikov}(2014)}]{Rubio:2014wta}%
  \BibitemOpen
  \bibfield  {author} {\bibinfo {author} {\bibfnamefont {J.}~\bibnamefont
  {Rubio}}\ and\ \bibinfo {author} {\bibfnamefont {M.}~\bibnamefont
  {Shaposhnikov}},\ }\href {\doibase 10.1103/PhysRevD.90.027307} {\bibfield
  {journal} {\bibinfo  {journal} {Phys. Rev. D}\ }\textbf {\bibinfo {volume}
  {90}},\ \bibinfo {pages} {027307} (\bibinfo {year} {2014})},\ \Eprint
  {http://arxiv.org/abs/1406.5182} {arXiv:1406.5182 [hep-ph]} \BibitemShut
  {NoStop}%
\bibitem [{\citenamefont {Casas}\ \emph {et~al.}(2019)\citenamefont {Casas},
  \citenamefont {Karananas}, \citenamefont {Pauly},\ and\ \citenamefont
  {Rubio}}]{Casas:2018fum}%
  \BibitemOpen
  \bibfield  {author} {\bibinfo {author} {\bibfnamefont {S.}~\bibnamefont
  {Casas}}, \bibinfo {author} {\bibfnamefont {G.~K.}\ \bibnamefont
  {Karananas}}, \bibinfo {author} {\bibfnamefont {M.}~\bibnamefont {Pauly}}, \
  and\ \bibinfo {author} {\bibfnamefont {J.}~\bibnamefont {Rubio}},\ }\href
  {\doibase 10.1103/PhysRevD.99.063512} {\bibfield  {journal} {\bibinfo
  {journal} {Phys. Rev. D}\ }\textbf {\bibinfo {volume} {99}},\ \bibinfo
  {pages} {063512} (\bibinfo {year} {2019})},\ \Eprint
  {http://arxiv.org/abs/1811.05984} {arXiv:1811.05984 [astro-ph.CO]}
  \BibitemShut {NoStop}%
\bibitem [{\citenamefont {Herrero-Valea}\ \emph {et~al.}(2019)\citenamefont
  {Herrero-Valea}, \citenamefont {Timiryasov},\ and\ \citenamefont
  {Tokareva}}]{Herrero-Valea:2019hde}%
  \BibitemOpen
  \bibfield  {author} {\bibinfo {author} {\bibfnamefont {M.}~\bibnamefont
  {Herrero-Valea}}, \bibinfo {author} {\bibfnamefont {I.}~\bibnamefont
  {Timiryasov}}, \ and\ \bibinfo {author} {\bibfnamefont {A.}~\bibnamefont
  {Tokareva}},\ }\href {\doibase 10.1088/1475-7516/2019/11/042} {\bibfield
  {journal} {\bibinfo  {journal} {JCAP}\ }\textbf {\bibinfo {volume} {11}},\
  \bibinfo {pages} {042} (\bibinfo {year} {2019})},\ \Eprint
  {http://arxiv.org/abs/1905.08816} {arXiv:1905.08816 [hep-ph]} \BibitemShut
  {NoStop}%
\bibitem [{\citenamefont {Rubio}(2020)}]{Rubio:2020zht}%
  \BibitemOpen
  \bibfield  {author} {\bibinfo {author} {\bibfnamefont {J.}~\bibnamefont
  {Rubio}},\ }in\ \href@noop {} {\emph {\bibinfo {booktitle} {{19th Hellenic
  School and Workshops on Elementary Particle Physics and Gravity}}}}\
  (\bibinfo {year} {2020})\ \Eprint {http://arxiv.org/abs/2004.00039}
  {arXiv:2004.00039 [gr-qc]} \BibitemShut {NoStop}%
\bibitem [{\citenamefont {Gurdogan}\ and\ \citenamefont
  {Kazakov}(2016)}]{Gurdogan:2015csr}%
  \BibitemOpen
  \bibfield  {author} {\bibinfo {author} {\bibfnamefont {O.}~\bibnamefont
  {Gurdogan}}\ and\ \bibinfo {author} {\bibfnamefont {V.}~\bibnamefont
  {Kazakov}},\ }\href {\doibase 10.1103/PhysRevLett.117.201602} {\bibfield
  {journal} {\bibinfo  {journal} {Phys. Rev. Lett.}\ }\textbf {\bibinfo
  {volume} {117}},\ \bibinfo {pages} {201602} (\bibinfo {year} {2016})},\
  \bibinfo {note} {[Addendum: Phys.Rev.Lett. 117, 259903 (2016)]},\ \Eprint
  {http://arxiv.org/abs/1512.06704} {arXiv:1512.06704 [hep-th]} \BibitemShut
  {NoStop}%
\bibitem [{\citenamefont {Karananas}\ \emph {et~al.}(2019)\citenamefont
  {Karananas}, \citenamefont {Kazakov},\ and\ \citenamefont
  {Shaposhnikov}}]{Karananas:2019fox}%
  \BibitemOpen
  \bibfield  {author} {\bibinfo {author} {\bibfnamefont {G.~K.}\ \bibnamefont
  {Karananas}}, \bibinfo {author} {\bibfnamefont {V.}~\bibnamefont {Kazakov}},
  \ and\ \bibinfo {author} {\bibfnamefont {M.}~\bibnamefont {Shaposhnikov}},\
  }\href@noop {} {\  (\bibinfo {year} {2019})},\ \Eprint
  {http://arxiv.org/abs/1908.04302} {arXiv:1908.04302 [hep-th]} \BibitemShut
  {NoStop}%
\bibitem [{\citenamefont {Englert}\ \emph {et~al.}(1976)\citenamefont
  {Englert}, \citenamefont {Truffin},\ and\ \citenamefont
  {Gastmans}}]{Englert:1976ep}%
  \BibitemOpen
  \bibfield  {author} {\bibinfo {author} {\bibfnamefont {F.}~\bibnamefont
  {Englert}}, \bibinfo {author} {\bibfnamefont {C.}~\bibnamefont {Truffin}}, \
  and\ \bibinfo {author} {\bibfnamefont {R.}~\bibnamefont {Gastmans}},\ }\href
  {\doibase 10.1016/0550-3213(76)90406-5} {\bibfield  {journal} {\bibinfo
  {journal} {Nucl. Phys. B}\ }\textbf {\bibinfo {volume} {117}},\ \bibinfo
  {pages} {407} (\bibinfo {year} {1976})}\BibitemShut {NoStop}%
\bibitem [{\citenamefont {Shaposhnikov}\ and\ \citenamefont
  {Tkachev}(2009)}]{Shaposhnikov:2008ar}%
  \BibitemOpen
  \bibfield  {author} {\bibinfo {author} {\bibfnamefont {M.~E.}\ \bibnamefont
  {Shaposhnikov}}\ and\ \bibinfo {author} {\bibfnamefont {I.~I.}\ \bibnamefont
  {Tkachev}},\ }\href {\doibase 10.1016/j.physletb.2009.04.040} {\bibfield
  {journal} {\bibinfo  {journal} {Phys. Lett. B}\ }\textbf {\bibinfo {volume}
  {675}},\ \bibinfo {pages} {403} (\bibinfo {year} {2009})},\ \Eprint
  {http://arxiv.org/abs/0811.1967} {arXiv:0811.1967 [hep-th]} \BibitemShut
  {NoStop}%
\bibitem [{\citenamefont {Armillis}\ \emph {et~al.}(2013)\citenamefont
  {Armillis}, \citenamefont {Monin},\ and\ \citenamefont
  {Shaposhnikov}}]{Armillis:2013wya}%
  \BibitemOpen
  \bibfield  {author} {\bibinfo {author} {\bibfnamefont {R.}~\bibnamefont
  {Armillis}}, \bibinfo {author} {\bibfnamefont {A.}~\bibnamefont {Monin}}, \
  and\ \bibinfo {author} {\bibfnamefont {M.}~\bibnamefont {Shaposhnikov}},\
  }\href {\doibase 10.1007/JHEP10(2013)030} {\bibfield  {journal} {\bibinfo
  {journal} {JHEP}\ }\textbf {\bibinfo {volume} {10}},\ \bibinfo {pages} {030}
  (\bibinfo {year} {2013})},\ \Eprint {http://arxiv.org/abs/1302.5619}
  {arXiv:1302.5619 [hep-th]} \BibitemShut {NoStop}%
\bibitem [{\citenamefont {Gretsch}\ and\ \citenamefont
  {Monin}(2015)}]{Gretsch:2013ooa}%
  \BibitemOpen
  \bibfield  {author} {\bibinfo {author} {\bibfnamefont {F.}~\bibnamefont
  {Gretsch}}\ and\ \bibinfo {author} {\bibfnamefont {A.}~\bibnamefont
  {Monin}},\ }\href {\doibase 10.1103/PhysRevD.92.045036} {\bibfield  {journal}
  {\bibinfo  {journal} {Phys. Rev. D}\ }\textbf {\bibinfo {volume} {92}},\
  \bibinfo {pages} {045036} (\bibinfo {year} {2015})},\ \Eprint
  {http://arxiv.org/abs/1308.3863} {arXiv:1308.3863 [hep-th]} \BibitemShut
  {NoStop}%
\bibitem [{\citenamefont {Almeida}\ \emph {et~al.}(2019)\citenamefont
  {Almeida}, \citenamefont {Bernal}, \citenamefont {Rubio},\ and\ \citenamefont
  {Tenkanen}}]{Almeida:2018oid}%
  \BibitemOpen
  \bibfield  {author} {\bibinfo {author} {\bibfnamefont {J.~P.~B.}\
  \bibnamefont {Almeida}}, \bibinfo {author} {\bibfnamefont {N.}~\bibnamefont
  {Bernal}}, \bibinfo {author} {\bibfnamefont {J.}~\bibnamefont {Rubio}}, \
  and\ \bibinfo {author} {\bibfnamefont {T.}~\bibnamefont {Tenkanen}},\ }\href
  {\doibase 10.1088/1475-7516/2019/03/012} {\bibfield  {journal} {\bibinfo
  {journal} {JCAP}\ }\textbf {\bibinfo {volume} {1903}},\ \bibinfo {pages}
  {012} (\bibinfo {year} {2019})},\ \Eprint {http://arxiv.org/abs/1811.09640}
  {arXiv:1811.09640 [hep-ph]} \BibitemShut {NoStop}%
\bibitem [{\citenamefont {Blas}\ \emph {et~al.}(2011)\citenamefont {Blas},
  \citenamefont {Shaposhnikov},\ and\ \citenamefont
  {Zenhausern}}]{Blas:2011ac}%
  \BibitemOpen
  \bibfield  {author} {\bibinfo {author} {\bibfnamefont {D.}~\bibnamefont
  {Blas}}, \bibinfo {author} {\bibfnamefont {M.}~\bibnamefont {Shaposhnikov}},
  \ and\ \bibinfo {author} {\bibfnamefont {D.}~\bibnamefont {Zenhausern}},\
  }\href {\doibase 10.1103/PhysRevD.84.044001} {\bibfield  {journal} {\bibinfo
  {journal} {Phys. Rev. D}\ }\textbf {\bibinfo {volume} {84}},\ \bibinfo
  {pages} {044001} (\bibinfo {year} {2011})},\ \Eprint
  {http://arxiv.org/abs/1104.1392} {arXiv:1104.1392 [hep-th]} \BibitemShut
  {NoStop}%
\bibitem [{\citenamefont {Broy}\ \emph {et~al.}(2015)\citenamefont {Broy},
  \citenamefont {Galante}, \citenamefont {Roest},\ and\ \citenamefont
  {Westphal}}]{Broy:2015qna}%
  \BibitemOpen
  \bibfield  {author} {\bibinfo {author} {\bibfnamefont {B.~J.}\ \bibnamefont
  {Broy}}, \bibinfo {author} {\bibfnamefont {M.}~\bibnamefont {Galante}},
  \bibinfo {author} {\bibfnamefont {D.}~\bibnamefont {Roest}}, \ and\ \bibinfo
  {author} {\bibfnamefont {A.}~\bibnamefont {Westphal}},\ }\href {\doibase
  10.1007/JHEP12(2015)149} {\bibfield  {journal} {\bibinfo  {journal} {JHEP}\
  }\textbf {\bibinfo {volume} {12}},\ \bibinfo {pages} {149} (\bibinfo {year}
  {2015})},\ \Eprint {http://arxiv.org/abs/1507.02277} {arXiv:1507.02277
  [hep-th]} \BibitemShut {NoStop}%
\end{thebibliography}%

\end{document}